\pdfoutput=1
\documentclass[aps,reprint,showpacs,twocolumn, superscriptaddress,preprintnumbers, amsmath, epsfig,floatfix]{revtex4-1}
\usepackage{graphicx}
\usepackage{amsmath}
\usepackage{amssymb}
\usepackage{amsfonts}
\usepackage{dcolumn}
\usepackage{dsfont}
\usepackage{latexsym}
\usepackage{rotating}
\usepackage{color}
\usepackage{latexsym}
\usepackage{bbm}
\usepackage{subfigure}
\usepackage{float}
\usepackage{epsfig}
\usepackage{psfrag}
\usepackage{natbib}
\usepackage{bm}
\usepackage{amsthm}
\usepackage{eucal}
\usepackage{mathrsfs}
\usepackage{url}
\usepackage{braket}
\usepackage{mathtools}
\usepackage{amsmath,amssymb}
\usepackage{epsfig}
\usepackage{amsmath,amssymb}
\usepackage{hyperref}

\usepackage{color}

\usepackage{hyperref}
\hypersetup{
colorlinks=true,final=true,
        linkcolor=blue,
        citecolor=blue,
        filecolor=blue,
        urlcolor=blue,
}

\begin{document}

\title{Origin and tuning of room-temperature multiferroicity in Fe doped BaTiO$_3$}

\author{Pratap Pal}
\affiliation{Department of Physics, Indian Institute of Technology Kharagpur, W.B. 721302, India}
\author{Krishna Rudrapal}
\affiliation{Advanced Technology Development Centre, Indian Institute of Technology Kharagpur, W.B. 721302, India}
\author{Sudipta Mahana}
\affiliation{Rajdhani College, Baramunda square, Bhubaneswar 751003, India}
\author{Satish Yadav}
\affiliation{UGC-DAE Consortium for Scientific Research, University Campus, Khandwa Road, Indore 452001, India}
\author{Tapas Paramanik}
\affiliation{Department of Physics, Indian Institute of Technology Kharagpur, W.B. 721302, India}
\affiliation{Department of Physics, School of Sciences, National Institute of Technology Andhra Pradesh, Tadepalligudem- 534101, India}
\author{Kiran Singh $\dagger$}
\affiliation{UGC-DAE Consortium for Scientific Research, University Campus, Khandwa Road, Indore 452001, India}
\author{Dinesh Topwal}
\affiliation{Institute of Physics, Sachivalaya Marg, Bhubaneswar 751005, India}
\author{Ayan Roy Chaudhuri}
\affiliation{Advanced Technology Development Centre, Indian Institute of Technology Kharagpur, W.B. 721302, India}
\affiliation{Materials Science Centre, Indian Institute of Technology Kharagpur, W.B. 721302, India}
\author{Debraj Choudhury}
\email{debraj@phy.iitkgp.ac.in}
\email{$^\dagger$Presently at Department of Physics, Dr. B. R. Ambedkar National Institute of Technology, Jalandhar 144011, India}
\affiliation{Department of Physics, Indian Institute of Technology Kharagpur, W.B. 721302, India}


\begin{abstract}

 Simultaneous co-existence of room-temperature ferromagnetism and ferroelectricity in Fe doped BaTiO$_3$ (BTO) is intriguing, as such Fe doping into tetragonal BTO, a room-temperature ferroelectric, results in the stabilization of its hexagonal polymorph which is ferroelectric only below $\sim {80}$K. Here, we investigate its origin and show that Fe doped BTO has a mixed-phase room-temperature multiferroicity, where the ferromagnetism comes from the majority hexagonal phase and a minority tetragonal phase gives rise to the observed weak ferroelectricity. In order to achieve majority tetragonal phase (responsible for room-temperature ferroelectricity) in Fe doped BTO, we investigate the role of different parameters which primarily control the paraelectric hexagonal phase stability over the ferroelectric tetragonal one and identify three major factors namely, the effect of ionic size, Jahn-Teller (J-T) distortions and oxygen-vacancies, to be primarily responsible. The effect of ionic size which can be qualitatively represented using the Goldschmidt's tolerance factor seems to be the major dictating factor for the hexagonal phase stability. The understanding of these factors not only enables us to control them but also, achieve suitable co-doped BTO compound with enhanced room-temperature multiferroic properties.
\end{abstract}

\maketitle

\section{Introduction}

The realization of room-temperature magnetoelectric-multiferroic materials which are simultaneously ferromagnetic and ferroelectric is extremely important due to their potential applicability in logic and information storage devices. However, it is extremely challenging to achieve so and there exists very few such examples \cite{JFScott2013,TKatayama2017,NAHill2000}. Recently, turning BaTiO$_3$, a room temperature proper ferroelectric, into a potential multiferroic material has been intensively studied \cite{RVKMangalam2009,SRamakanth2014,ARaeliarijaona2017}. Interestingly, it has been found that multiferroicity can emerge in BTO nano-particles in which ferromagnetism comes from the surface states due to oxygen-vacancies and the ferroelectricity from the core \cite{RVKMangalam2009,SRamakanth2014}. Later, it was also shown that in addition to oxygen, Ti or Ba vacancies can also induce ferromagnetism in BTO \cite{ARaeliarijaona2017}. But as the induced magnetism by such vacancies is very small (few memu/gm) at room temperature \cite{RVKMangalam2009}, there were efforts towards the realization of multiferroicity in BTO by other routes like transition metal ion ( Mn, Fe or Co) doping at Ti sites \cite{YHLin2009,RMaier2001,HKChandra2013,LBLuo2009}. In this regard, Fe doped BTO, which exhibits the simultaneous coexistence of room-temperature ferromagnetism and ferroelectricity \cite{BXu2009,ARani2016}, attracted a lot of attentions. However, such Fe doping induced transformation of the ferroelectric tetragonal BTO (t-BTO) into its hexagonal polymorph (h-BTO), which is though paraelectric at room temperature \cite{SRay2008,ESawaguchi1985}. The origin of ferromagnetism which likely arises due to the 90$^0$ superexchange interaction between two Fe ions at Ti2 sites in the BTO hexagonal matrix \cite{SRay2008}, can be enhanced with increasing oxygen-vacancy content \cite{TChakraborty2011}. However, the tunability of ferromagnetism by creating oxygen vacancies can be detrimental to the insulating ferroelectric properties of BTO \cite{ASagdeo2018}. Thus, it is important to find other possible ways to tune the ferromagnetic properties of such Fe doped BTO system. In this regard, another major issue is to recover back the ferroelectric tetragonal phase from paraelectric hexagonal Fe doped BTO, such that we can have optimized multiferroic response from such system. Therefore, it is important to identify the key controlling parameters that facilitate the hexagonal phase formation, such that hexagonal to tetragonal structural phase tuning becomes possible.


\begin{figure}[h!]
	
	\vspace*{-0.00 in}
	\hspace*{-0.06in} \scalebox{0.25}{\includegraphics{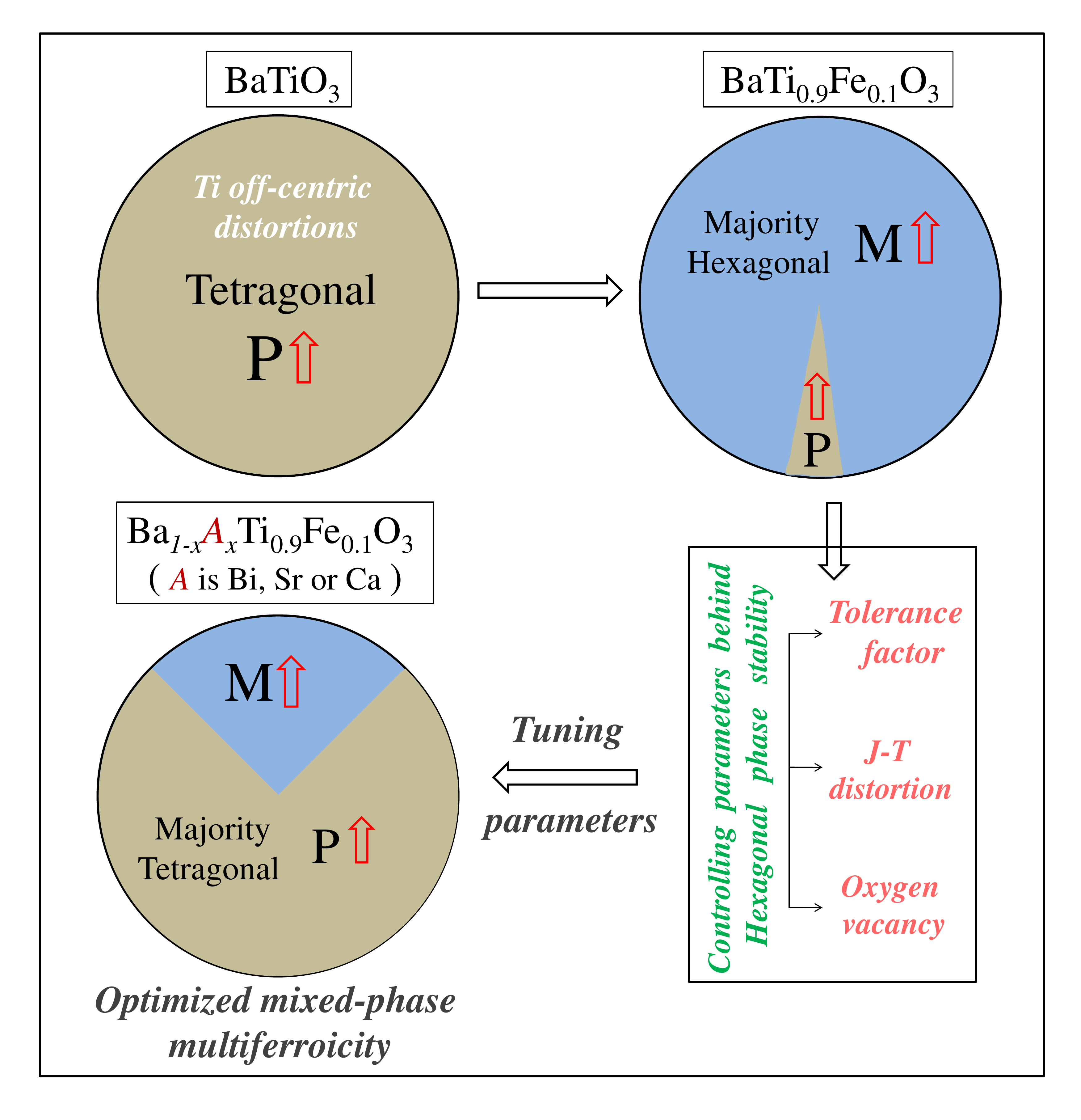}}
	\vspace*{-0.15in}\caption{Schematic representation of room-temperature mixed phase multiferroicity in Fe doped BTO. Majority hexagonal phase, which can be controlled by tuning either Goldschmidt's tolerance factor, J-T distortion of dopant ion or oxygen-vacancy content, contributes to the ferromagnetism (M$\uparrow$) of Fe doped BTO. The observed weak ferroelectricity (P$\uparrow$) of Fe doped BTO arises from trace quantity of minority tetragonal phase. Identification and tuning of the control parameters which stabilize the paraelectric hexagonal phase over ferroelectric tetragonal phase in Fe doped BTO lead to optimized mixed-phase multiferroic response from this system.}\label{MixedPhaseMultiferroicity}
	
\end{figure}

  The magnetic ground state of Fe doped BTO which is ferromagnetic at room-temperature, remains contentious, as both paramagnetism \cite{XKWei2012} and ferromagnetism \cite{SRay2008} have been proposed to exist at low temperature. Interestingly, a recent experimental result suggests a high-temperature ferromagnetic to low-temperature paramagnetic phase transition in Fe doped BTO \cite{XKWei2012}. Like magnetism, the origin of room-temperature ferroelectric polarization of Fe doped BTO is also intriguing, since undoped h-BTO is paraelectric at room-temperature \cite{ESawaguchi1985}. Also, both simultaneous ferromagnetic-paraelectric \cite{SRay2008} and ferromagnetic-ferroelectric \cite{BXu2009,ARani2016} behaviors have been observed  in Fe doped BTO, thus raising the question on the origin of the observed ferroelectricity and its reproducibility. It is interesting to note that ferroelectric instability can also be triggered due to pseudo-Jahn Teller effect (PJTE) of the doped Fe$^{3+}$ (d$^5$) ions in the hexagonal Fe doped BTO, as recently shown in LaFeO$_3$ \cite{LWeston2016}. Therefore, it is very critical to understand the origin of ferroelectricity and magnetic state such that the tuning of its room-temperature multiferroic properties becomes feasible.

With these goals in mind, we examine high quality polycrystalline BTO and various Ba$_{1-{\it{x}}}${\it{A}}$_{\it{x}}$Ti$_{0.9}$Fe$_{0.1}$O$_{3}$ (where {\it{A}} is Bi, Sr or Ca and {\it{x}} denotes the atomic concentration of doped ion at the dopant site) and Ba$_{0.9}$Bi$_{0.10}$Ti$_{0.9}$Mn$_{0.1}$O$_{3}$ compounds. Here, in this paper, we show that only Fe doped BTO is actually a mixed phase room-temperature multiferroics, where a majority hexagonal phase is ferromagnetic and a minority tetragonal phase gives rise to the observed weak ferroelectricity (see Fig.\ref{MixedPhaseMultiferroicity}). Subsequently, we show that along with J-T distortions and oxygen-vacancies, the contribution of Goldschmidt's tolerance factor on the hexagonal phase stabilization is extremely important and by manipulating these parameters, we are able to enhance the room-temperature multiferroicity.

\section{Experimental Methods}

All samples are prepared in phase-pure polycrystalline form via solid state reactions. The properly ground mixture of individual oxide elements taken in stoichiometric amounts, is first annealed at 1050$^0$C for 12 hour and then at 1250$^0$C for 12 hour to get the final product. Room-temperature powder XRD and Raman spectroscopic measurements are carried out using Cu-K$\alpha$ source and a 514 nm line of Ar$^+$ laser as an excitation source, respectively. The adopted nomenclatures for all samples are listed in table-\ref{Sampledetails}. The ferroelectric PUND (Positive-Up-Negative-Down) measurements \cite{JTEvans2011} are conducted in Radiant P-E loop tracer and the dc magnetization measurements are carried out using a Quantum Design SQUID magnetometer. The temperature dependent pyroelectric current measurements are carried out using Keithley 6517B electrometer. Samples are first cooled down to 10K from room temperature under the presence of applied electric field of 400V/mm. Then, the electric field is switched off and the electrodes are short-circuited for sufficient time to get rid of any residual surface charge effects. Sample is then heated to room temperature at a rate of 10 K/min and we record temperature dependent pyroelectric current. X-ray Photoelectron Spectroscopy (XPS) studies were carried out using an in-house PHI 5000 Versaprobe-II spectrometer and X-ray Absorption Near Edge Spectroscopy (XANES) investigations are performed at P-65 beamline in PETRA-III synchrotron source, DESY, Hamburg, Germany.

\begin{table}[h!]
	\centering
	\caption{All sample notations and chemical formulas along with their respective hexagonal volume phase fractions, as obtained from Rietveld refinement of their room temperature powder XRD patterns. The sample notation, for example, BB05TFO (Ba$_{0.95}$Bi$_{0.05}$Ti$_{0.9}$Fe$_{0.1}$O$_3$) means Bi and Fe are simultaneously doped at Ba and Ti sites of BTO by 5 and 10 atomic\% respectively, where the notation ``05" denotes the atomic\% of Bi doping at Ba sites. }
	\label{Sampledetails}
	\vspace*{0.5cm}
	\begin{tabular}{|c|c|c|c|}
		\hline
		
		Sample & Desired chemical & Hexagonal phase  \\
		notations & formula & fraction (\%)  \\
		\hline
		BTFO & BaTi$_{0.9}$Fe$_{0.1}$O$_3$ & 98.8  \\
		BTFO\_Q & BTFO quenched in LN$_2$ & 99.3 \\
		BB02TFO & Ba$_{0.98}$Bi$_{0.02}$Ti$_{0.9}$Fe$_{0.1}$O$_3$ & 38.2 \\
		BB05TFO & Ba$_{0.95}$Bi$_{0.05}$Ti$_{0.9}$Fe$_{0.1}$O$_3$ & 18.6  \\
		BB10TFO &Ba$_{0.9}$Bi$_{0.1}$Ti$_{0.9}$Fe$_{0.1}$O$_3$ & 0  \\
		BS05TFO & Ba$_{0.95}$Sr$_{0.05}$Ti$_{0.9}$Fe$_{0.1}$O$_3$ & 24.6  \\
		BC05TFO & Ba$_{0.95}$Ca$_{0.05}$Ti$_{0.9}$Fe$_{0.1}$O$_3$ & 23.2  \\
		BB10TMO & Ba$_{0.9}$Bi$_{0.1}$Ti$_{0.9}$Mn$_{0.1}$O$_3$ & 13.3  \\
		\hline
	\end{tabular}
	
\end{table}


\begin{figure}[h!]
	
	\vspace*{-0 in}
	\hspace*{-0.2in} \scalebox{0.4}{\includegraphics{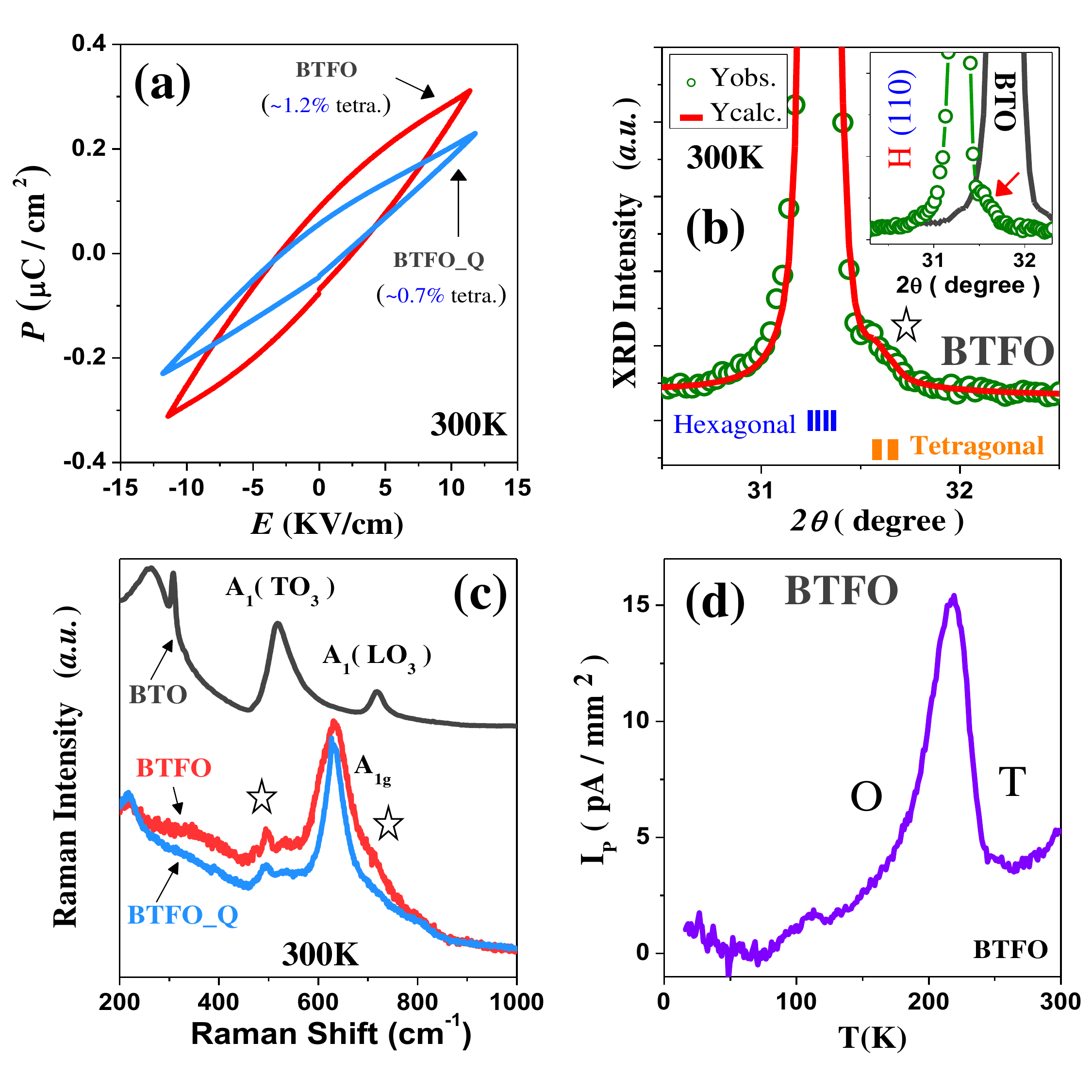}}
	\vspace*{-0.25in}\caption{(Color online) (a) Comparison of room-temperature P-E loops of BTFO and BTFO\_Q, (b) signature of tetragonal phase from both raw XRD pattern and Rietveld refinement, (c) Room-temperature Raman spectra of BTFO, BTFO\_Q and pure BTO (star marks show the presence of tetragonal phase), and (d) Temperature-dependence of pyroelectric-current for BTFO which shows ferroelectric-orthorhombic (O) to ferroelectric-tetragonal (T) phase transition at ∼218K during heating.}\label{FerroelectricOrigin}
	
\end{figure}


\begin{figure}[h!]
	
	\vspace*{-0 in}
	\hspace*{-0.15in} \scalebox{0.4}{\includegraphics{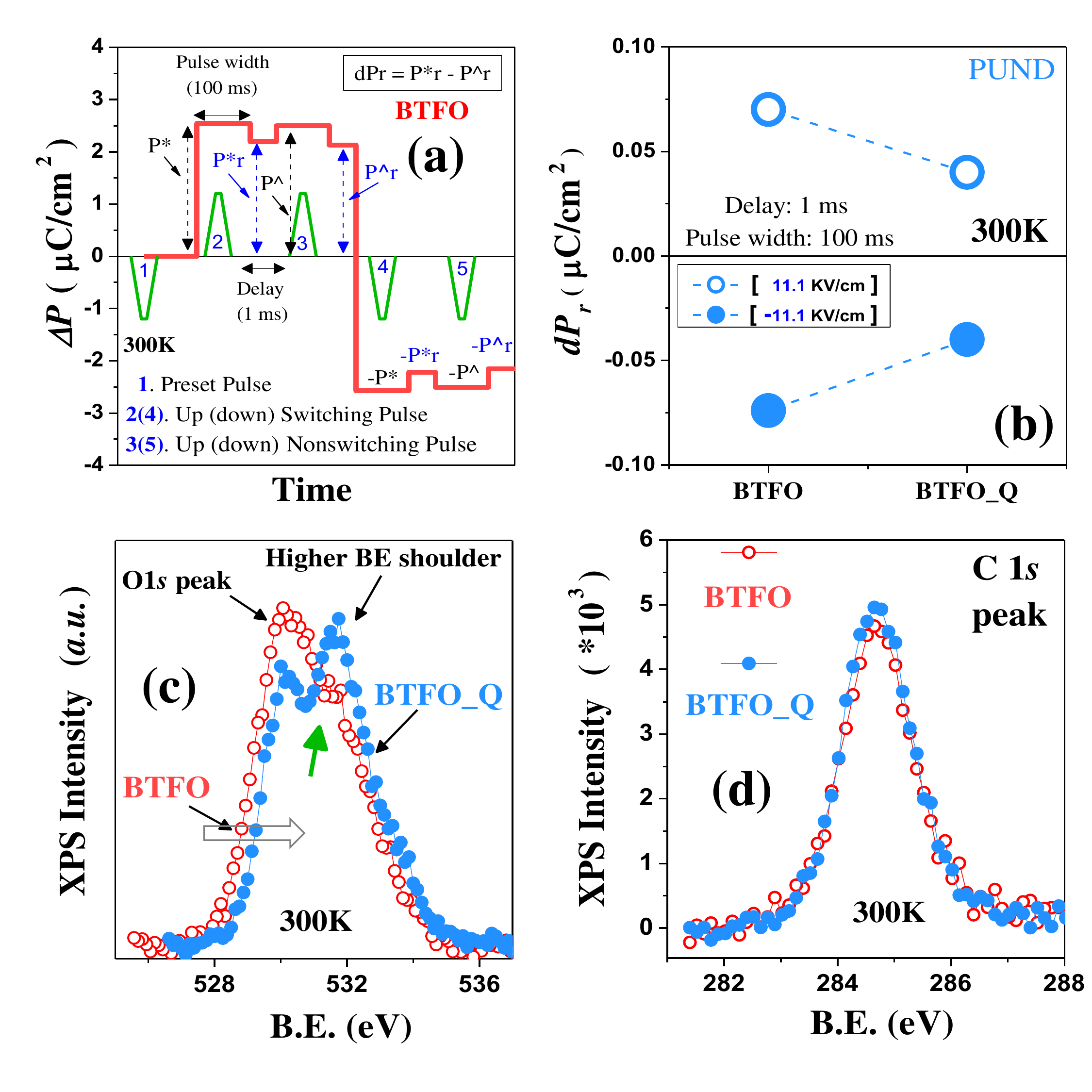}}
	\vspace*{-0.3in}\caption{(Color online) (a) Room-temperature PUND data (red line) of BTFO, where the applied voltage pulses (green ones) are schematically shown, (b) variation and switching of the room-temperature remanent polarization of BTFO and BTFO\_Q as obtained from PUND measurement, (c) shows corresponding  room-temperature O-1{\it{s}} XPS spectra and (d) shows their room-temperature C-1{\it{s}} XPS spectra.}\label{PUND-XPS}
	
\end{figure}

\section{Results and Discussions}

\subsection{Fe-doped BaTiO$_3$: a mixed-phase multiferroics}

 The room-temperature powder XRD pattern of BTFO, shown in Fig.\ref{BTFOXRD} (supplementary section), nicely matches to that of BTO hexagonal polymorph with the space group {\it{P6$_3$/mmc }} \cite{RBottcher2008}. Though, h-BTO is paraelectric at room-temperature \cite{ESawaguchi1985}, surprisingly, we find some finite ferroelectric polarization in hexagonal BTFO, as shown in Fig.\ref{FerroelectricOrigin}(a). Importantly, the detailed XRD spectrum of BTFO exhibits a small, but broad shoulder just after the hexagonal (110) peak, which can be accounted for by considering an additional t-BTO phase (shown in the inset to Fig.\ref{FerroelectricOrigin}(b)). The corresponding Rietveld refinement, taking mixed hexagonal and tetragonal phases, is shown in Fig.\ref{FerroelectricOrigin}(b). Raman spectroscopy on BTFO supports this observation of mixed-phase, where clear signature of tetragonal Raman modes A$_1$(TO$_3$) and A$_1$(LO$_3$) (shown by star marks) \cite{CHPerry1965}, along with the majority hexagonal A$_{1g}$ mode \cite{HYamaguchi1987} are observed, as shown in Fig.\ref{FerroelectricOrigin}(c). These observations, thus clearly indicate that BTFO is actually a mixed phase compound in which majority is hexagonal (~98.8\%, obtained from Rietveld refinement) and the rest is weak tetragonal. Further, we employ pyroelectric measurement which is very sensitive to ferroelectric phase transition. Fig.\ref{FerroelectricOrigin}(d) clearly shows a phase transition at $\sim$ 218K for BTFO. As BTFO has both hexagonal and tetragonal phases and hexagonal phase being ferroelectric only below $\sim$ 80K, such transition in pyrocurrent is an indication of a ferroelectric phase transition in the minority tetragonal phase from Ti off-centric displacements \cite{RECohen1992}. For such a scenario, tuning of the tetragonal phase fraction should also lead to changes in the observed room-temperature ferroelectric polarization value. BTFO\_Q, which expectedly contains more oxygen-vacancies than BTFO, as engineered through quenching in liquid N$_2$ during synthesis \cite{GWang2019}, is found to contain a lower tetragonal phase fraction, as clearly seen through the weaker tetragonal Raman peaks in Fig.\ref{FerroelectricOrigin}(c). Following the above conjecture, BTFO\_Q indeed possesses a lower room-temperature polarization value when compared to BTFO, as shown in Fig.\ref{FerroelectricOrigin}(a). This is further highlighted using a comparison of the true remanent ferroelectric polarization values between BTFO and BTFO\_Q, as extracted using the PUND technique, which is a well established method to extract out the intrinsic ferroelectric part (remanent contribution) from other non-ferroelectric (non-remanent) contributions \cite{JTEvans2011,JFScott1989,YSChai2012}. Fig.\ref{PUND-XPS}(a) displays room-temperature PUND polarization data of BTFO, where the preset, switching and non-switching applied voltage pulses are schematically shown (see also Fig.\ref{PUND} and Fig.\ref{PUND-all} in the supplementary section). The negative pulse-1 is necessary to preset the material. The application of the positive switching pulse-2, thereby, induces switching of both the ferroelectric component along with switching of the linear non-ferroelectric (non-remanent) contributions. As the ferroelectric dipoles are already aligned and do not relax during the intermittent delay time, the following positive pulse-3 cannot induce any further switching of the ferroelectric component and only induces further polarization of the non-ferroelectric (non-remanent) components. Thus, the intrinsic ferroelectric remanent polarization (dPr) can be readily obtained by subtracting the polarization (P$^\wedge$r) value, obtained after application of pulse-3, from the polarization value (P$^\star$r), obtained after the application of pulse-2. A comparison of the dPr values for BTFO and BTFO\_Q, obtained using the above method, are shown in Fig.\ref{PUND-XPS}(a). We indeed see the presence and switching of remanent polarization (dPr) for both BTFO and BTFO\_Q as shown in Fig.\ref{PUND-XPS}(b). The PUND data clearly show a smaller remanent polarization value for BTFO\_Q in comparison to BTFO (similar trend as observed in Fig.\ref{FerroelectricOrigin}(a)), which can be expected only if the ferroelectric polarization originates from the tetragonal phase. Now, to understand the presence of oxygen vacancies, we have carried out room temperature XPS measurements. It is well known that the higher binding-energy shoulder of O-1{\it{s}} peak arises from defect sites, like oxygen-vacancies and adsorbed hydrocarbons \cite{SJaiswar2017}. The presence of identical weightages of C-1{\it{s}} spectra for both BTFO and BTFO\_Q, as seen in the Fig.\ref{PUND-XPS}(d), show the presence of similar amounts of adsorbed hydrocarbons on the surface of these samples. However, the higher binding-energy shoulder of O-1{\it{s}} peak for BTFO\_Q has significantly larger intensity as compared to BTFO, as shown in Fig.\ref{PUND-XPS}(c), thereby, clearly bringing out the presence of more oxygen-vacancies in BTFO\_Q sample \cite{SJaiswar2017,WLi2014}. This is also supported by the movement of O-1{\it{s}} peak to higher binding-energy in BTFO\_Q as compared to BTFO (see Fig.\ref{PUND-XPS}(c)) \cite{GLiu2009}. Interestingly, while stoichiometric BTFO should contain only nominal Fe$^{4+}$ (the ground state in this case is primarily a mixture of {\it{d}}$^4$ and {\it{d}}$^5$$\underbar{L}$ states with $\underbar{L}$ representing an O 2{\it{p}} hole \cite{AEBocquet1992}) ions, presence of oxygen-vacancies is correlated with the conversion of some nominal Fe$^{4+}$ ions into Fe$^{3+}$ state due to charge neutrality \cite{RBottcher2008,HaMNguyen2011}. Thus, the presence of smaller room-temperature ferroelectric polarization value in BTFO\_Q, which contains more Fe$^{3+}$ ion content than BTFO, also rules out PJTE as the mechanism for the observed ferroelectric polarization of BTFO.


\begin{figure}[t]
	
	\vspace*{0 in}
	\hspace*{-0.1in} \scalebox{0.4}{\includegraphics{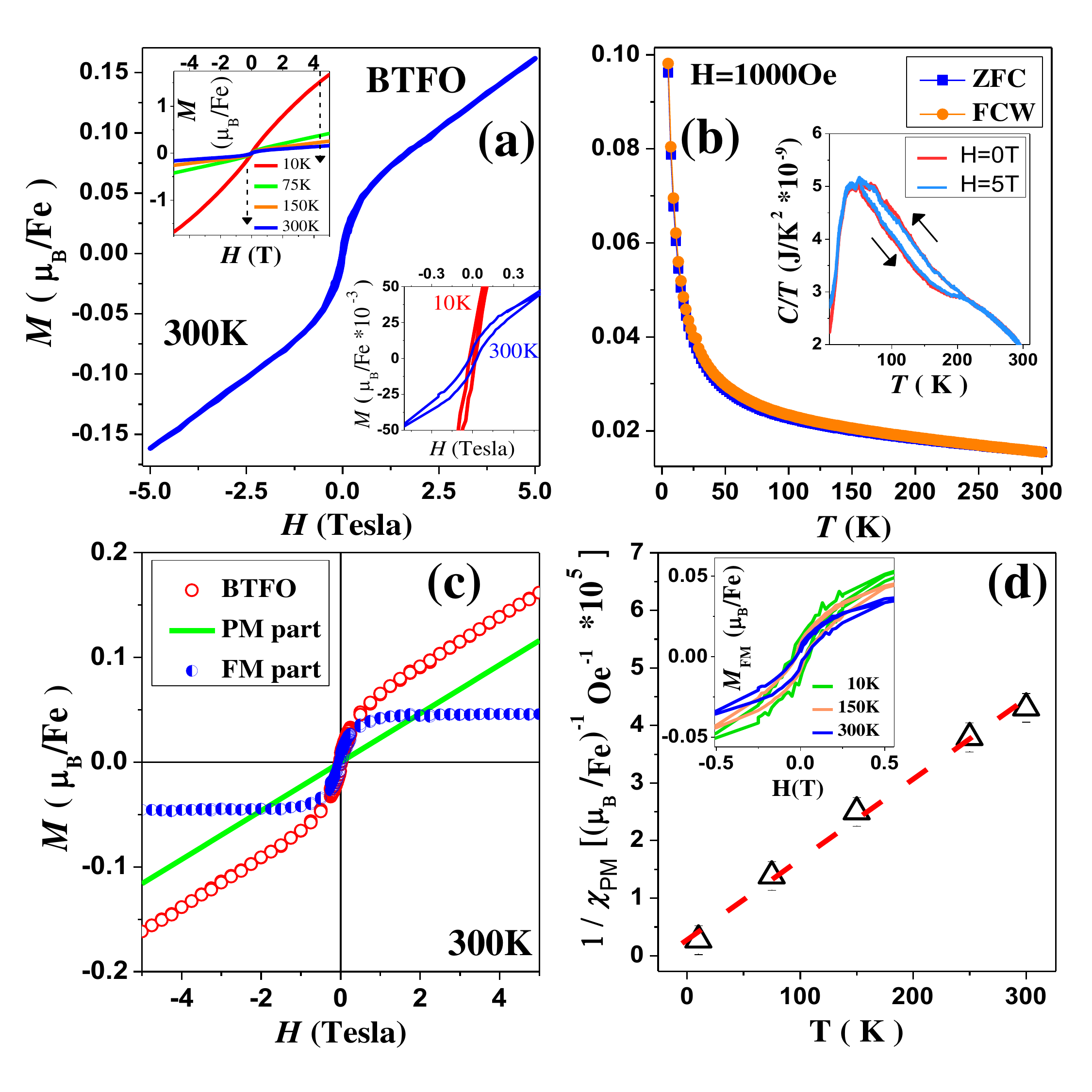}}
	\vspace*{-0.25in}\caption{(Color online) (a) Isothermal M-H loops of BTFO and (b) variation of magnetization (M) with temperature (T). Inset to (b) shows the heat capacity vs. T plot, both in absence and presence of magnetic field of 5 Tesla. (c) Decomposition of room-temperature M-H plot into paramagnetic and ferromagnetic parts and (d) displays the linear behavior of the paramagnetic inverse susceptibility with temperature. Inset to (d) shows the existence of ferromagnetic loops at all temperatures.}\label{Magnetic}
	
\end{figure}

\subsection{Coexistence of ferromagnetism and paramagnetism in Fe-doped BaTiO$_3$}

To understand the magnetic ground state, we have performed isothermal M-H measurements from room-temperature down to 10K, as shown in Fig.\ref{Magnetic}(a). At room-temperature, there is a clear ferromagnetic M-H loop for BTFO, which is understood to arise from ferromagnetic exchange interactions in the hexagonal phase \cite{SRay2008,INApostolova2013}. At 10K, the M-H loop, however, looks like a Brillouin function, as expected for paramagnetic order, as shown in Fig.\ref{Magnetic}(a). Such a behavior has previously been attributed to a room-temperature ferromagnetic to low-temperature paramagnetic phase transition for BTFO \cite{XKWei2012}, which is quite unusual, and, thus, we investigated it further. On carefully analyzing the 10K M-H loop near low-field region, we do find existence of a clear hysteresis loop, also suggesting the presence of ferromagnetic order at low-temperature. Also, although a clear M-H hysteresis loop is observed at 300K, unlike a ferromagnet, it does not saturate and continues to rise linearly in the high field region. Further, no clear ferromagnetic to low-temperature paramagnetic phase transition is observed in M vs.T plot, as shown in Fig.\ref{Magnetic}(b) or even in dM/dT vs.T plot. The absence of any magnetic transition is also supported by the heat capacity data, which do not show any significant changes in presence of magnetic field (H=5Tesla) (as shown in the inset to Fig.\ref{Magnetic}(b)), thereby, suggesting that the transitions observed in the intermediate temperature window are of structural origin. The structural origin of the heat-capacity anomalies are also evident from the temperature-dependent dielectric constant data (see Fig.\ref{BTFO-DC} of supplementary section). A structural transition can indeed be expected from the high-temperature hexagonal to low-temperature orthorhombic phase around similar temperature \cite{SRay2008}. It is also important to note that often such dilute magnetic semiconductor (DMS) systems are associated with both paramagnetic and ferromagnetic signals \cite{SRay2008,SIAndronenko2015}, as seems to be the case for BTFO. We, thus, fit the high field magnetization data at 300K with a linear equation and the obtained slope gives us the estimate of the paramagnetic contribution (M=$\chi$H). On performing similar analyses at all temperatures, we find the expected linear (1/$\chi$ -T) dependence of the extracted paramagnetic contributions, as shown in Fig.\ref{Magnetic}(d). Simultaneously, the presence of clear M-H loops (obtained by subtracting the paramagnetic linear contribution) at all temperatures from 300K down to 10K, suggest that ferromagnetic order exists down to the lowest-temperature. Thus, our data clearly suggest co-existence of both ferromagnetic and paramagnetic regions in BTFO at all temperatures.


\begin{figure}[t]
	
	\vspace*{-0.14 in}
	\hspace*{-0.15in} \scalebox{0.4}{\includegraphics{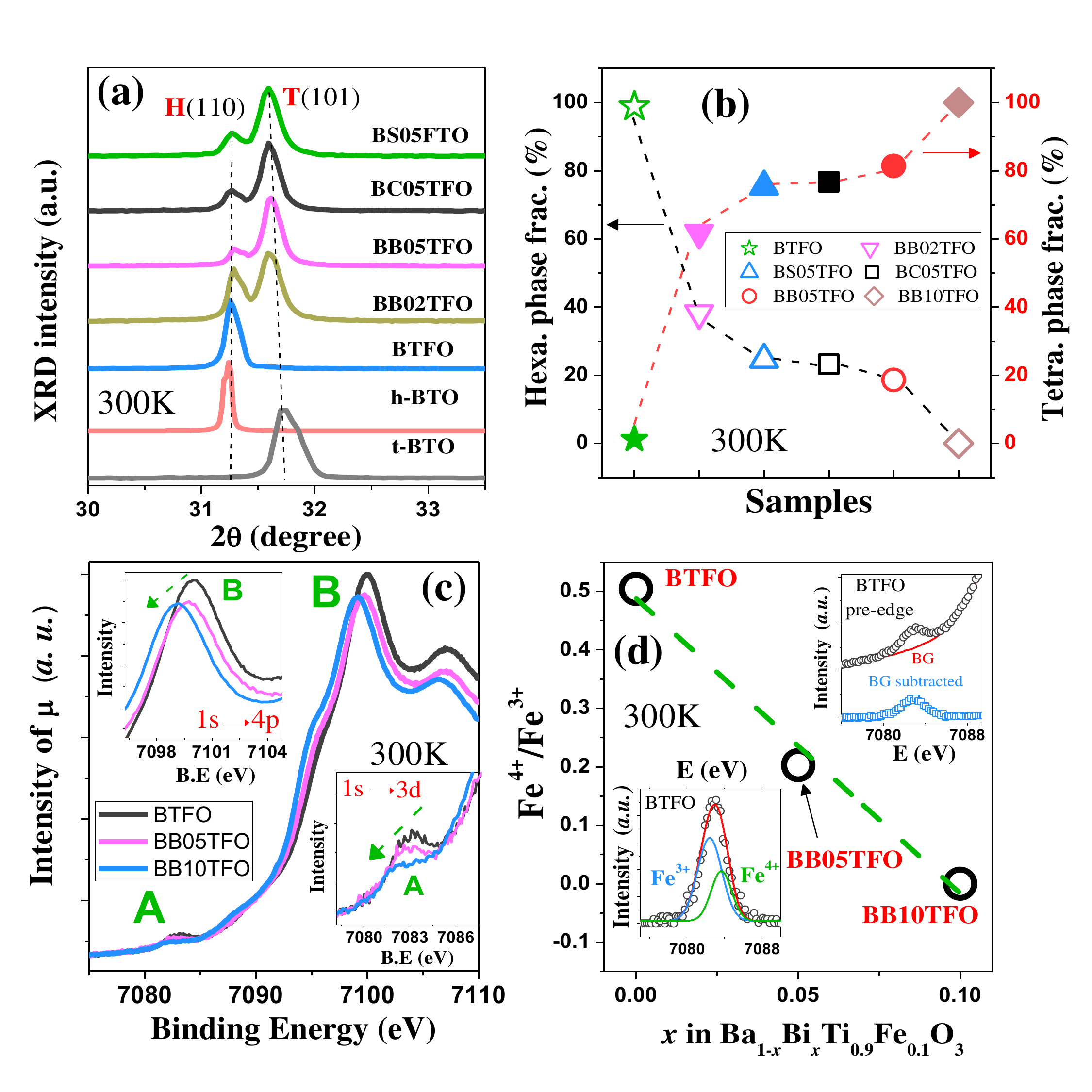}}
	\vspace*{-0.33in}\caption{(Color online) (a) Room-temperature XRD spectra (only h-BTO reference XRD is taken from ICSD database) of the samples. (b) Variation of hexagonal phase (left y-axis) and tetragonal phase (right y-axis) fractions (\%) for different sample [(Hexa. + Tetra) phase fractions = 100\%]. (c) Room-temperature Fe K-edge XANES spectra of BTFO, BB05TFO and BB10TFO. The top and bottom insets show the evolution of main edge and pre-edge regions with doping. (d) Quantification of nominal Fe$^{4+}$ to Fe$^{3+}$ ratio, through background (BG) subtraction (top inset) and deconvolution of the pre-edge region into Fe$^{3+}$ and nominal Fe$^{4+}$ contributions (bottom inset).}\label{HexagonalPhase}
	
\end{figure}


\begin{figure}[t]
	
	\vspace*{-0 in}
	\hspace*{-0.15in} \scalebox{0.4}{\includegraphics{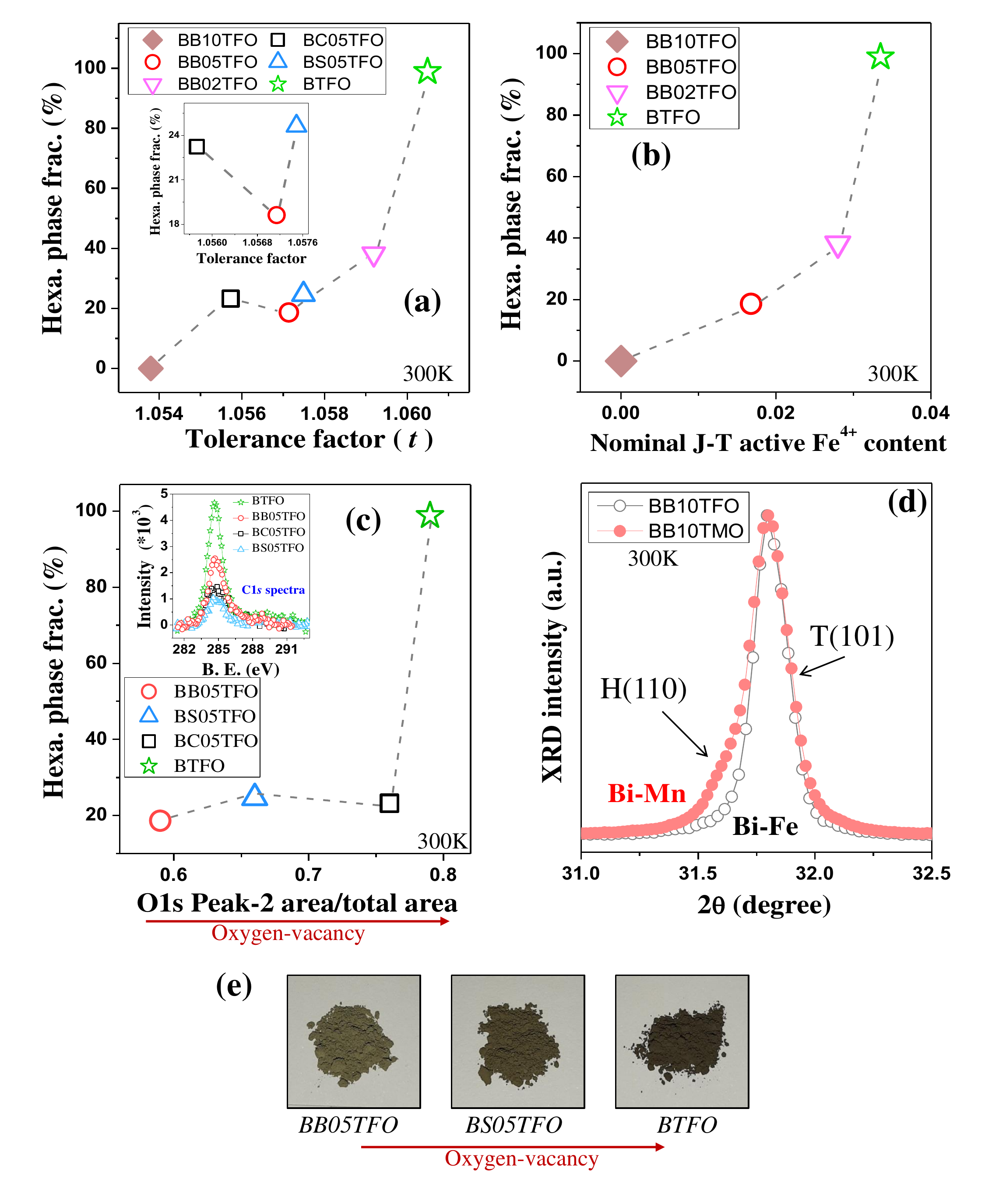}}
	\vspace*{-0.3in}\caption{(Color online) Variation of hexagonal phase fractions (\%) [100 - tetragonal phase fraction(\%)] of the samples with their corresponding (a) Goldschmidt's tolerance factor, (b) the nominal J-T active Fe$^{4+}$ content present, (c) O-1{\it{s}} higher binding-energy peak (peak-2) area divided by the total area, which corresponds to the oxygen-vacancy in the system. The inset to (c) displays their C-1{\it{s}} spectra. (d) Room-temperature XRD spectra of BB10TFO and BB10TMO, where clear presence of significant amount of hexagonal phase in BB10TMO, indicates the role of J-T distortion due to Mn$^{3+}$ (3d$^4$) towards hexagonal phase stabilization and (e) shows variation of sample color, dictating the change in oxygen-vacancy.}\label{HexaControlParameters}
	
\end{figure}


\begin{figure}[t]
	
	\vspace*{-0.05 in}
	\hspace*{-0.18in} \scalebox{0.4}{\includegraphics{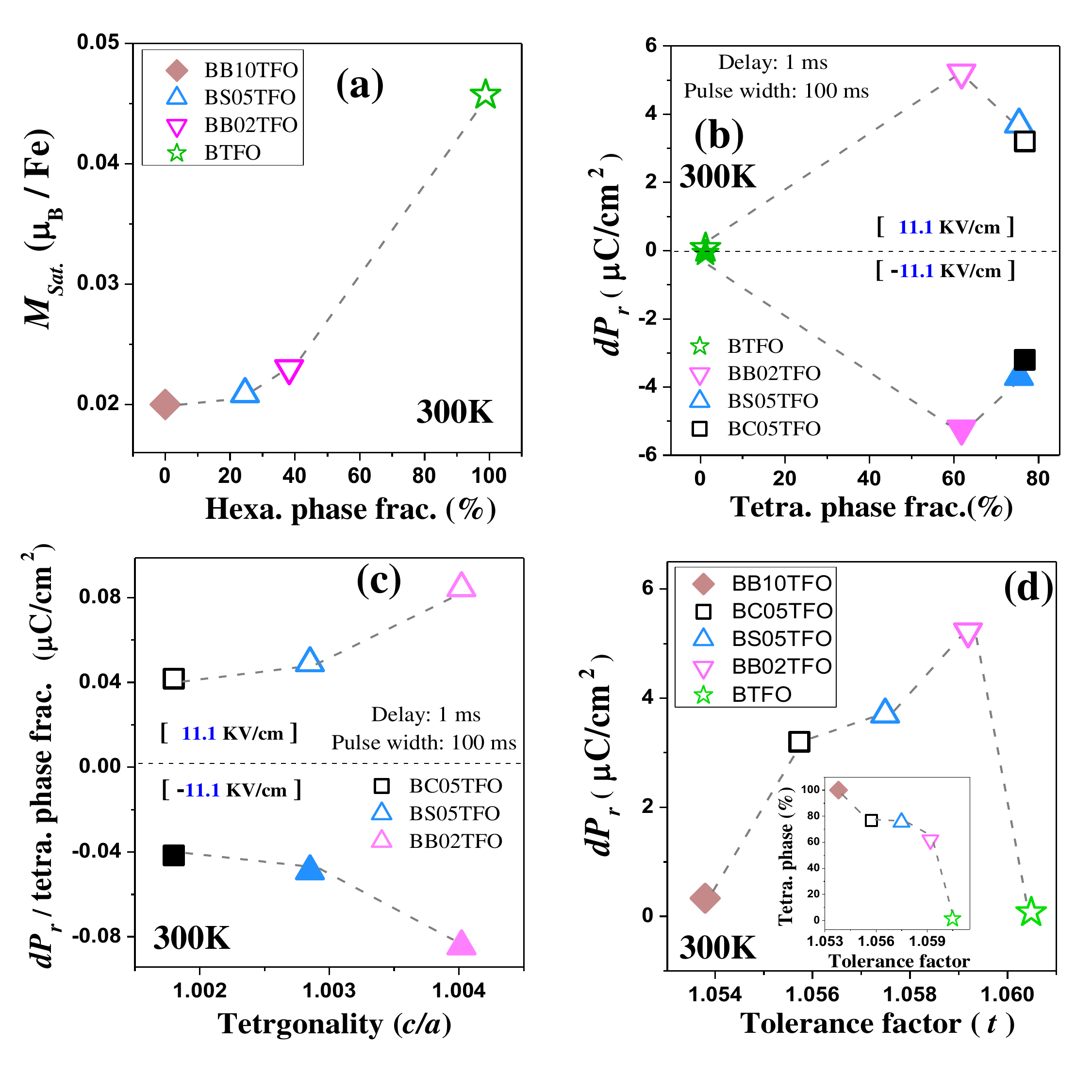}}
	\vspace*{-0.28in}\caption{(Color online) (a) Variation of room-temperature saturation magnetization (M$_S$) of the extracted ferromagnetic part with the corresponding sample hexagonal phase fraction, (b) variation as well as switching of the room-temperature remanent polarization {\color{red}{(dP$_r$)}} as obtained from PUND measurements with the corresponding tetragonal phase fraction (\%), (c) variation of remanent polarization per unit tetragonal phase fraction (\%)  with the corresponding sample tetragonality ({\it{c/a}}) and (d) displays the change in remanent polarization for samples with their corresponding Goldschmidt's tolerance factor, where its inset shows how tetragonal phase fraction (\%) varies with their respective tolerance factor.}\label{TuningProperties}
	
\end{figure}

\subsection{The role of controlling parameters behind hexagonal phase stability in Fe-doped BaTiO$_3$ }

 We have, thus, elucidated that mixed phase multiferroicity exists in Fe doped BTO, with weak ferroelectricity arising from the minority tetragonal phase. To enhance ferroelectricity in doped BTO, it is important to investigate the role of different parameters that play crucial part in the hexagonal phase stability in BTFO. For this purpose, we examine the BTFO, BTFO\_Q, BB02TFO, BB05TFO, BS05TFO, BC05TFO, BB10TFO and BB10TMO samples. Room-temperature XRD spectra are shown in Fig.\ref{HexagonalPhase}(a), where we can clearly see variations of hexagonal first-order peak intensity for different samples. The percentage of hexagonal and tetragonal phases present in the samples are determined from Rietveld refinement and are tabulated in Table-\ref{Sampledetails} and also shown in Fig.\ref{HexagonalPhase}(b) (for refinement details see Fig.\ref{AllRefinement}, Fig.\ref{Structure}, Fig.\ref{StructuralInformation} and Fig.\ref{Refinement details} in the supplementary section). The hexagonality which is here found to be maximum for only Fe doped BTO (BTFO), gets further enhanced in BTFO\_Q due to the presence of more oxygen-vacancies compared to BTFO. Such strong hexagonal phase stabilization over tetragonal BTO on Fe doping is known to be primarily dictated by Jahn-Teller distortion of nominally Fe$^{4+}$ ions and further secondary contribution comes from oxygen-vacancies \cite{RBottcher2008,HTLanghammer2000,SGCao2017}. As Ba is gradually replaced by Bi, Sr or Ca, the hexagonal phase fraction is found to get strongly suppressed along with a concomitant tetragonal BTO phase recovery, as shown in Fig.\ref{HexagonalPhase}(b) (see also Table-\ref{Sampledetails}). To understand the role of different controlling parameters on the hexagonal phase stability, we first consider the three compounds BB05TFO (Bi doped at Ba site), BS05TFO (Sr doped at Ba site) and BC05TFO (Ca doped at Ba site). The reduction of hexagonal phase fraction in BB05TFO can be understood using the frame work of previous studies, as stated earlier, since Bi$^{3+}$ substitution in place of Ba$^{2+}$ reduces oxygen-vacancies as well as J-T active nominally Fe$^{4+}$ content to Fe$^{3+}$ (3d$^5$) to maintain charge neutrality. Such observations are further validated from the systematic trend of hexagonal phase tuning from 98.8\% in case of BTFO (where only Fe is doped) to 0\% for BB10TFO (where Bi and Fe are doped in equal amounts) through 38.2\% in BB02TFO and 18.6\% in BB05TFO (see Fig.\ref{HexagonalPhase}(b) and Table-\ref{Sampledetails}). However, the sharp reduction of hexagonal phase by significant amounts on Sr$^{2+}$ or Ca$^{2+}$ substitutions in BS05TFO and BC05TFO is quite surprising and cannot be understood using the earlier framework, as doping with Sr$^{2+}$ and Ca$^{2+}$, being isovalent to Ba$^{2+}$ (only difference is their ionic sizes), is neither expected to vary the oxygen-vacancies nor the J-T active ion content. Thus, a holistic understanding of the role of ionic size effects, quantified through the Goldschmidt’s tolerance factor, which provides an effective way of investigating phase stabilities of perovskite oxides \cite{ZLi2016}, along with oxygen-vacancies and J-T distortions to the hexagonal phase stabilization becomes crucial.

 Since the ionic sizes depend on the ion-valency, we have carried out XANES measurements to investigate the Fe-ion valency. Ti was found to remain in 4+ state in all these samples, as seen by our XPS studies (see Fig.\ref{Ti2PXPS} of supplementary section). The normalized Fe K-edge XANES spectra of BTFO, BB05TFO and BB10TFO are shown in Fig.\ref{HexagonalPhase}(c). We can subdivide the spectra into two regions, the pre-edge region ({\it{A}}: 1{\it{s}} $\to 3{\it{d}}$ excitation) and the main edge region ({\it{B}}: 1{\it{s}} $\to 4{\it{p}}$ excitation) \cite{ERAluri2013}. Here, features {\it{A}} and {\it{B}} are observed to shift towards lower energy with increasing Bi(+3) doping, which is expected if the oxidation state of Fe steadily decreases from nominal Fe$^{4+}$ to Fe$^{3+}$ \cite{OHaas2009}. For BB10TFO with equal amount of Bi$^{3+}$ and Fe doping, Fe is expected to be in the Fe$^{3+}$ state. Hence, these XANES observations indicate the presence of nominal Fe$^{4+}$ to some extent in BTFO, which gets decreased in amount with Bi doping. To quantify nominal Fe$^{4+}$ to Fe$^{3+}$ ratio, we investigate the pre-edge peak region \cite{OHaas2009}. The tentative amounts of nominal Fe$^{4+}$ in these samples, as calculated from background subtraction and subsequent deconvolution of pre-edge peak into Fe$^{3+}$ and nominal Fe$^{4+}$ contributions are shown in Fig.\ref{HexagonalPhase}(d). From this calibration, now, we calculate Goldschmidt's tolerance factor for all samples.

 Fig.\ref{HexaControlParameters}(a) shows the variation of hexagonal phase fraction with the corresponding tolerance factor for various samples, the correspondence seems near direct; increase in the tolerance factor generally leads to an increase in the hexagonal phase fraction. It is interesting to note that hexagonality, which gets suppressed in BS05TFO (Ba$^{2+}$ is partially replaced by Sr$^{2+}$) compared to BTFO (Ba is not replaced), gets further suppressed in BC05TFO where Ba$^{2+}$ is replaced by Ca$^{2+}$ (having smaller ionic radii than Sr$^{2+}$) due to smaller tolerance factor of BC05TFO as compared to BS05TFO (as shown in the inset to Fig.\ref{HexaControlParameters}(a)). Such observations strongly indicate that the Goldschmidt’s tolerance factor or the ionic size effect play a crucial role behind the hexagonal phase stability. Against the above general trend, BB05TFO has smaller hexagonal phase fraction when compared to BC05TFO, though the tolerance factor of the former is more than that of the later (as shown in the inset to Fig. \ref{HexaControlParameters}(a)). To understand this, we need to also consider the effect of the other controlling parameters like oxygen-vacancies or J-T distortions, associated with nominal Fe$^{4+}$ ions, on the hexagonal phase stability.
 		
 In Fig.\ref{HexaControlParameters}(b) we note that as J-T active nominal Fe$^{4+}$ content (as obtained from room-temperature Fe-K edge XANES measurements) increases, hexagonal phase gets more stabilized. To elucidate it further, we have compared room-temperature XRD spectra of BB10TFO (Ba$_{0.9}$Bi$_{0.1}$Ti$_{0.9}$Fe$_{0.1}$O$_{3}$) with BB10TMO (Ba$_{0.9}$Bi$_{0.1}$Ti$_{0.9}$Mn$_{0.1}$O$_{3}$) in Fig.\ref{HexaControlParameters}(d), where the former (BB10TFO) contains only Jahn-Teller inactive Fe$^{3+}$ (3d$^5$) ion, and the later (BB10TMO) contains only Jahn-Teller active Mn$^{3+}$ (3d$^4$) ion. In agreement with the trend of increasing hexagonality with increasing content of Jahn-Teller active ion, it is found that while BB10TFO is completely tetragonal, BB10TMO has significant amount of hexagonal phase, though both of them have identical tolerance factor (Fe$^{3+}$ and Mn$^{3+}$ have same ionic radii).

 Next, to understand the role of oxygen-vacancies towards hexagonal phase stability, we have carried out room-temperature XPS studies on O-1{\it{s}} spectra (see Fig.\ref{Oxygenvacancy} of supplementary section). As discussed earlier, the higher binding-energy shoulder (peak-2) of O-1{\it{s}} spectra corresponds to oxygen vacancies and adsorbed hydrocarbons \cite{SJaiswar2017}. Upon fitting the O-1{\it{s}} spectra with two peaks, we plot the change in hexagonal phase with the corresponding O-1{\it{s}} peak-2 area (normalized by total area under O-1{\it{s}} peak region after Shirley background subtraction), as shown in Fig.\ref{HexaControlParameters}(c) (the inset shows the corresponding C-1{\it{s}} spectra which measures the relative adsorbed hydrocarbon content). The results show that there is a strong reduction of oxygen-vacancies on Bi$^{3+}$ doping for Ba$^{2+}$, which can be expected since increased Bi$^{3+}$ ion content naturally promotes the more stable Fe$^{3+}$ ion content without the need for any oxygen vacancies. Surprisingly, we also notice reduction of oxygen-vacancies in isovalent substituted BC05TFO and BS05TFO compounds. This observation can only be understood by considering the suppression of hexagonal phase in the Sr- and Ca-doped samples as it has been argued that oxygen vacancies tend to be hosted in the hexagonal phase \cite{DCSinclair1999,DLu2019} rather than the tetragonal phase. Further, such reduction of oxygen-vacancy content in the samples where the A site Ba ion is replaced by other ions (Bi, Sr or Ca) can also be visualized from sample color change as shown in Fig.\ref{HexaControlParameters}(e), {\color{blue}{where}} it seems that the darker the sample is, more is the oxygen-vacancy content. A comparison between Fig.\ref{HexaControlParameters}(a) and (c) shows that though BC05TFO has more oxygen vacancies than BS05TFO, it has lesser hexagonal phase fraction; which seems primarily due to smaller tolerance factor of the former compared to the later. Similarly, since BB05TFO has lower oxygen-vacancy and Jahn-Teller active ion content as compared to BC05TFO, the former has slightly smaller hexagonal phase fraction as compared to the later, as seen in the inset to Fig. 6(a).

 Therefore, from these discussions it is clear that along with oxygen-vacancy and J-T distortions, there is a significant role of ionic size towards hexagonal phase stabilization in such transition metal doped BaTiO$_3$. It is also extremely important to understand which of these controlling parameters plays a more dominant role to destabilize hexagonal phase, because then by controllably manipulating that parameter we can more effectively control the hexagonal to tetragonal structural phase ratio, which is critical for achieving better optimized room-temperature multiferroicity in such Fe doped BTO system. To highlight this, we consider two compounds BTFO and BB05TFO for which we see that hexagonal phase fraction is reduced by $\sim$80\% in the later compound compared to the former due to only $\sim$0.3\% change in tolerance factor, whereas the respective change in other controlling parameters (namely oxygen vacancy content and Jahn-Teller active ion content) is quite large. Such observation clearly indicates that the Goldschmidt’s tolerance factor plays the dictating role behind the hexagonal phase stability.

\begin{table}[h!]
	\centering
	\caption{All samples with their corresponding tetragonal volume phase fractions, room-temperature tetragonality ({\it{c/a}}) as obtained from Rietveld refinement and room temperature intrinsic remanent ferroelectric polarization as observed in PUND measurements.}
	\label{PolarizationTetragonality}
	\vspace*{0.5cm}
	\begin{tabular}{|c|c|c|c|}
		\hline
		
		Sample &  Tetragonal  & Tetragonality & Remanent   \\
		notations & phase fraction  & ({\it{c/a}}) &  polarization  \\
		& (\%) &  &  dPr ($\mu$C/cm$^2$) \\
		\hline
		&  &  &  \\
		BTFO & 1.2 & 1.0018 & 0.07 \\
		BTFO\_Q & 0.7 & 1.0016 & 0.04 \\
		BB02TFO & 61.8 & 1.0040 & 5.22 \\
		BB05TFO & 81.4 & 1.0037 & 4.3  \\
		BB10TFO & 100 & 1.0002 & 0.33 \\
		BS05TFO & 75.4 & 1.0029 & 3.7 \\
		BC05TFO & 76.8 & 1.0018 & 3.2 \\
		\hline
	\end{tabular}
	
\end{table}

\subsection{Tuning of room-temperature multiferroicity of Fe-doped BaTiO$_3$}

For the study of multiferroic properties, we have carried out room-temperature M-H measurements on BTFO, BB02TFO, BS05TFO and BB10TFO. The saturated magnetization (M$_S$) of the extracted ferromagnetic part for these samples along with the corresponding hexagonal phase fraction (\%) present in the samples are shown in Fig.\ref{TuningProperties}(a). Here, the change in M$_S$ closely reflects the change in the respective hexagonal volume phase fraction. It is already known that the ferromagnetism in Fe doped BTO arises due to 90$^0$ superexchange between two Fe2 ions at Ti2-sites in the hexagonal unit \cite{SRay2008}, hence the tuning of such hexagonal phase also leads to the tuning of corresponding ferromagnetic properties, as seen in Fig.\ref{TuningProperties}(a).
		
Subsequently, We have studied the intrinsic ferroelectric properties of these samples employing PUND technique. Fig.\ref{TuningProperties}(b) shows the variation as well as ferroelectric switching of remanent polarization (dPr) with the corresponding sample tetragonal phase fraction (\%). Here, we see that BB02TFO has maximum polarization (5.22 $\mu$C/cm$^2$) which is $\sim$74 times more than that of BTFO, which is quite remarkable, as the recovery of ferroelectricity in such Fe doped BTO system by such margin can be extremely helpful in the engineering of room temperature multiferroicity. Surprisingly, though BC05TFO has more tetragonal phase fractions than BS05TFO and BB02TFO, it shows lesser ferroelectric polarization as seen in Fig.7(b). To understand this, we have plotted ferroelectric polarization normalized to the tetragonal phase percentage present in the sample along with the corresponding tetragonality ({\it{c/a}}) as shown the Fig.\ref{TuningProperties}(c), where we find that the order of sample tetragonality is BB02TFO$>$BS05TFO$>$BC05TFO, which clearly reflects the origin of smaller polarization in BC05TFO. Thus, the sample tetragonality, evidently, seems to play a dominant role to the ferroelectric properties (see also Table-\ref{PolarizationTetragonality}).The strong dependency of the recovered ferroelectricity of all these compounds on the tetragonal phase fraction and tetragonality ({\it{c/a}}) indicates that the ferroelectricity originates from the Ti off-centric distortions \cite{RECohen1992}. We further note that all the investigated compounds have Goldschmidt’s tolerance factor $\geqslant$1.054, which is larger than the limit necessary to cause octahedral tilting \cite{JHLee2016,AMarthinsen2016}, as observed in orthorhombic CaMnO$_3$ \cite{JKlarbring2018}, thus, any role of such octahedral tilting to the observed ferroelectricity in our samples can be ruled out.

It is interesting to note that there is a range of Goldschmidt’s tolerance factor for which the sample shows optimized ferroelectric polarization, as seen in Fig.\ref{TuningProperties}(d). Such optimized range of tolerance factor for best ferroelectric Fe doped BTO compound arises because for higher tolerance factor the sample tends to have more hexagonality, whereas for smaller tolerance factors, the sample goes to cubic limit. Thus, the understanding of the origin of ferroelectricity and the controlling parameters for the hexagonal phase stability in Fe doped BTO, enables us to controllably tune the room-temperature mixed phase multiferroicity, where magnetism is attributed to the hexagonal phase and the ferroelectricity to the tetragonal phase.

\section{Conclusions}

 In summary, we have investigated the room-temperature multiferroicity in Fe doped BTO and have shown that it is of mixed phase origin, where the ferromagnetism is associated with the majority hexagonal phase and ferroelectricity to the minority tetragonal phase. We have also elucidated that ferromagnetic and paramagnetic phases coexist at all temperatures in BTFO. By examining various co-doped BTO compounds, we show that compared to J-T distortions and oxygen-vacancies, the Goldschmidt's tolerance factor or the ionic size effect has more decisive role on the stabilization of paraelectric hexagonal phase over the ferroelectric tetragonal one. By controlling these parameters, we successfully achieve tuning of the room-temperature mixed phase multiferroicity in engineered co-doped BTO compounds.

 \section{Acknowledgments}

We acknowledge the use of XPS under the DST-FIST facility in the Department of Physics, IIT Kharagpur for this work. P.P. would like to acknowledge the financial support from MHRD, India. D.C. would like to acknowledge SERB, DST, India (funding under project file no. ECR/2016/000019) and BRNS, DAE (funding through sanction number 37(3)/20/23/2016-BRNS) for financial support. D.T. would like to gratefully acknowledge financial support by DST provided with in the framework of the India@DESY collaboration.

\section{Supplementary information files}

\begin{figure*} 
	
	\centering \includegraphics{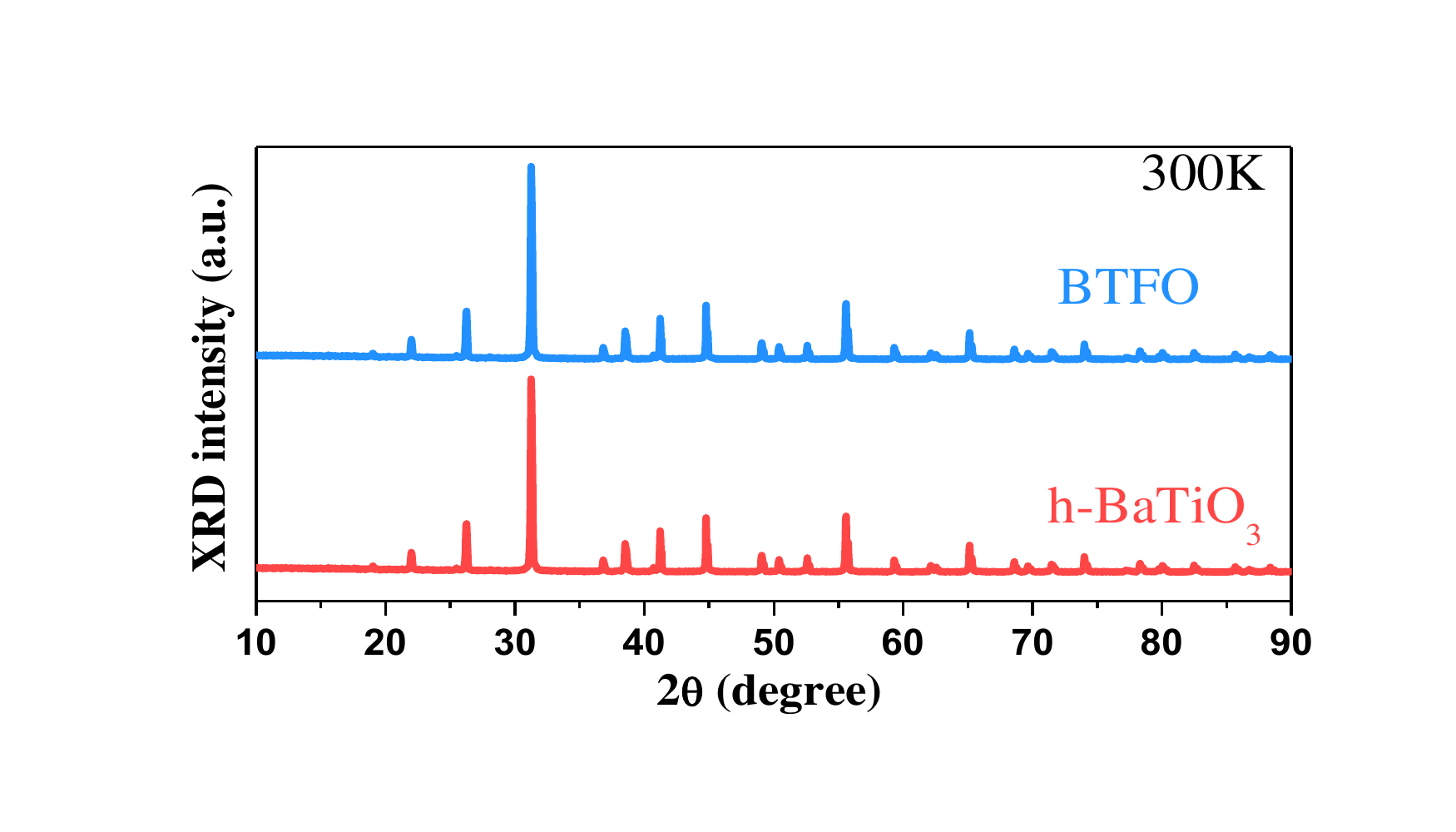}
	\caption{Room temperature XRD spectrum of BTFO (BaTi$_{0.9}$Fe$_{0.1}$O$_3$) matches to that of  BTO hexagonal polymorph with the space group {\it{P6$_3$/mmc }} (room temperature XRD spectrum of standard hexagonal BaTiO$_3$ is taken from ICSD database).}\label{BTFOXRD}
	
\end{figure*}

\subsection{The room temperature XRD spectrum of pure BaTi$_{0.9}$Fe$_{0.1}$O$_3$ (see Fig.\ref{BTFOXRD})}

\begin{figure*} 
	
	\centering \scalebox{0.38}{\includegraphics{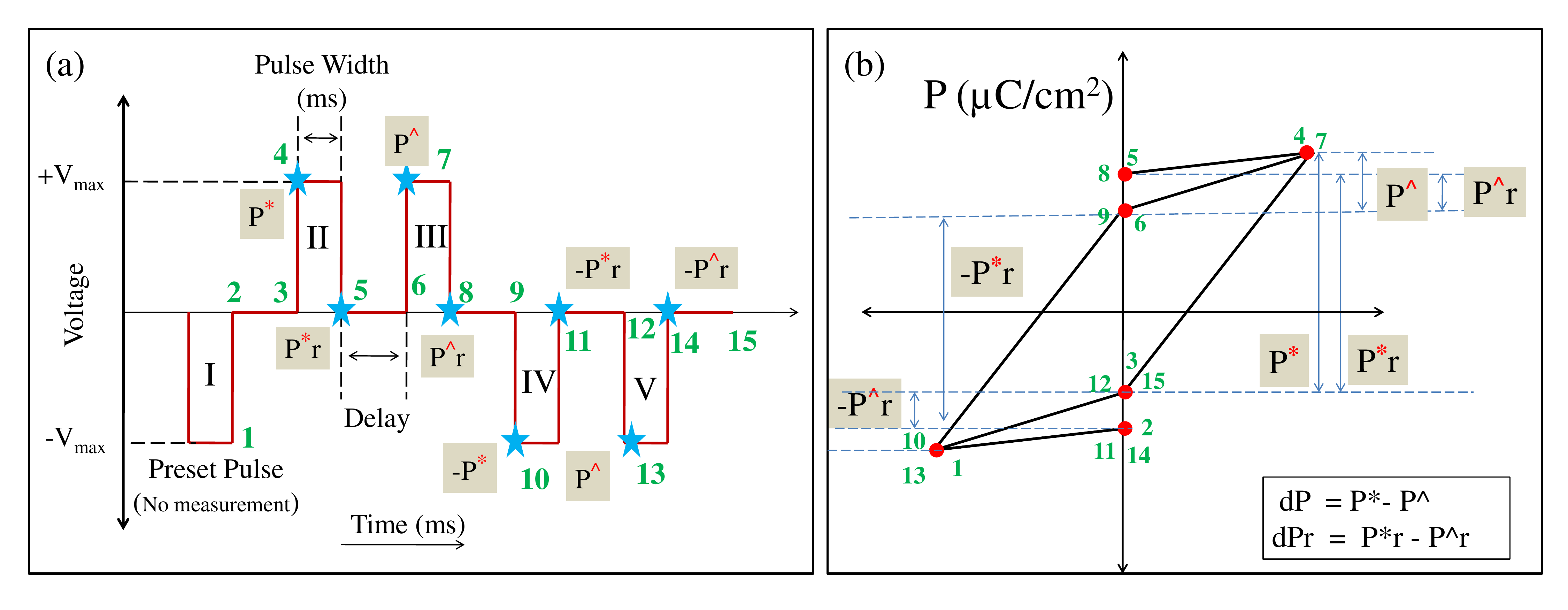}} 
	\caption{(a) Voltage pulse wave forms used in ferroelectric PUND measurement, and (b) corresponding polarization switching in response to the applied voltage pulses.The PUND technique is a standard ferroelectric memory characterization technique in which a series of five voltage pulses are applied. The pulse-I is used just to preset the sample without carrying out any measurement. The pulse-II switches the sign of the preset polarization and measures the amount of polarization switched. After the measurement, the voltage pulse is reduced to zero volts and the sample is kept through a sufficient delay time so that all non-remanent polarization dissipates. After this, pulse-III, same as pulse-II, is applied with the same measurements being made. In this case the sample is not switched, hence, the measured polarization will represent the non-Remanent content of the sample. Subsequently, the pulse-IV and pulse-V are just opposite to second and third pulses, switching the sample in the negative V$_{Max.}$
		direction. In this process, the remanent polarization is obtained as dPr = P$^\star$r - – P$^\wedge$r.
	}\label{PUND}
	
\end{figure*}

	\subsection{PUND mechanism (see Fig.\ref{PUND})}

\begin{figure*} 
	
	\centering \scalebox{0.7}{\includegraphics{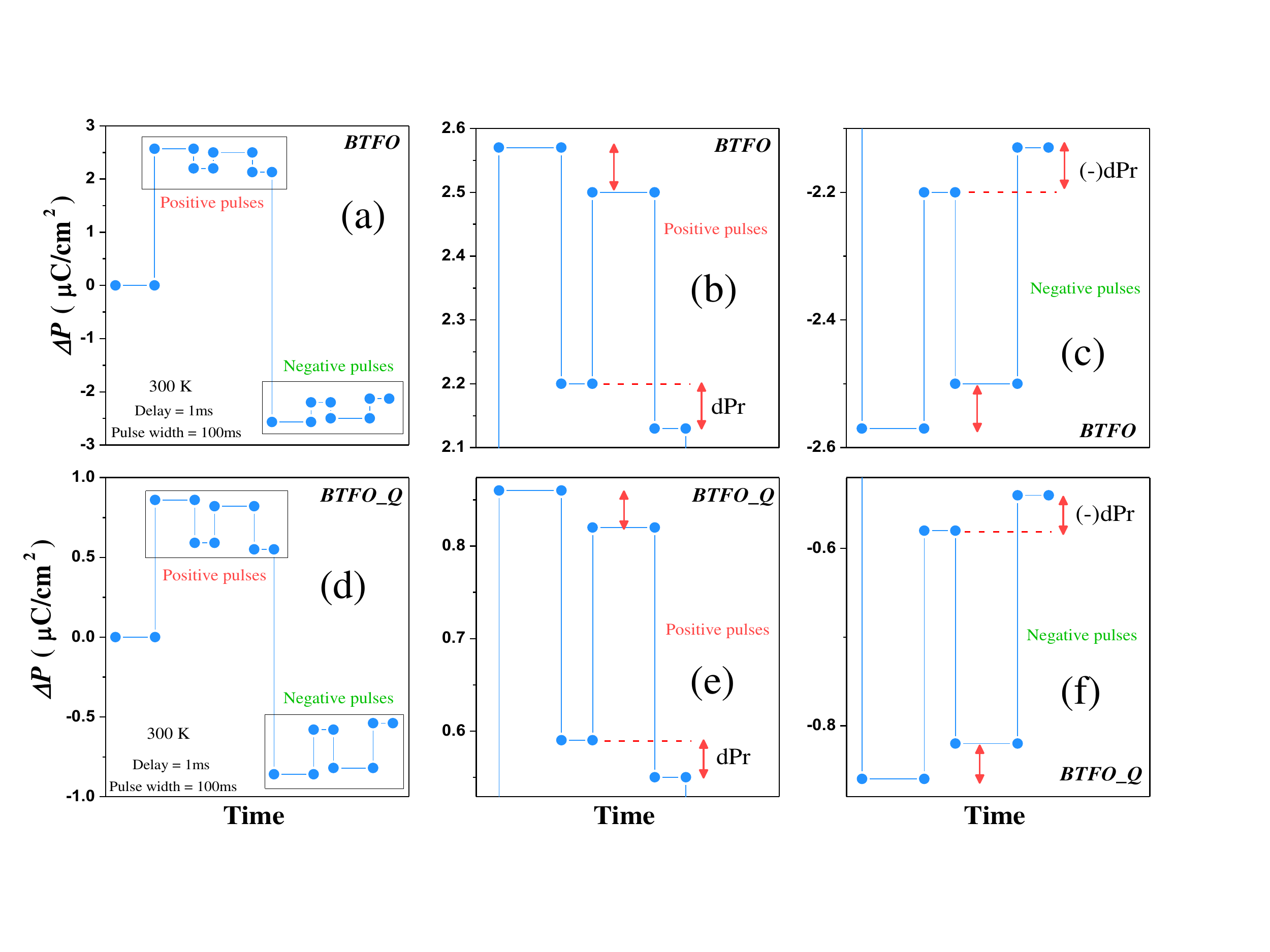}} 
	\caption{(a) and (d) display PUND results of polycrystalline BTFO (BaTi$_{0.9}$Fe$_{0.1}$O$_3$) and BTFO\_Q (a part of BTFO quenched in liquid nitrogen directly from 1250$^0$C) compounds respectively. (b), (c) and (e), (f) display close-up view of the polarization response obtained from positive and negative pulses applied on BTFO and BTFO\_Q respectively, which show presence of clear remanent polarization, establishing the existence of intrinsic ferroelectricity in these two compounds.}\label{PUND-all}
	
\end{figure*}

	\subsection{Intrinsic ferroelectricity as detected in PUND measurement (see Fig.\ref{PUND-all})}
	
\begin{figure*} 
	
	\centering \scalebox{0.85}{\includegraphics{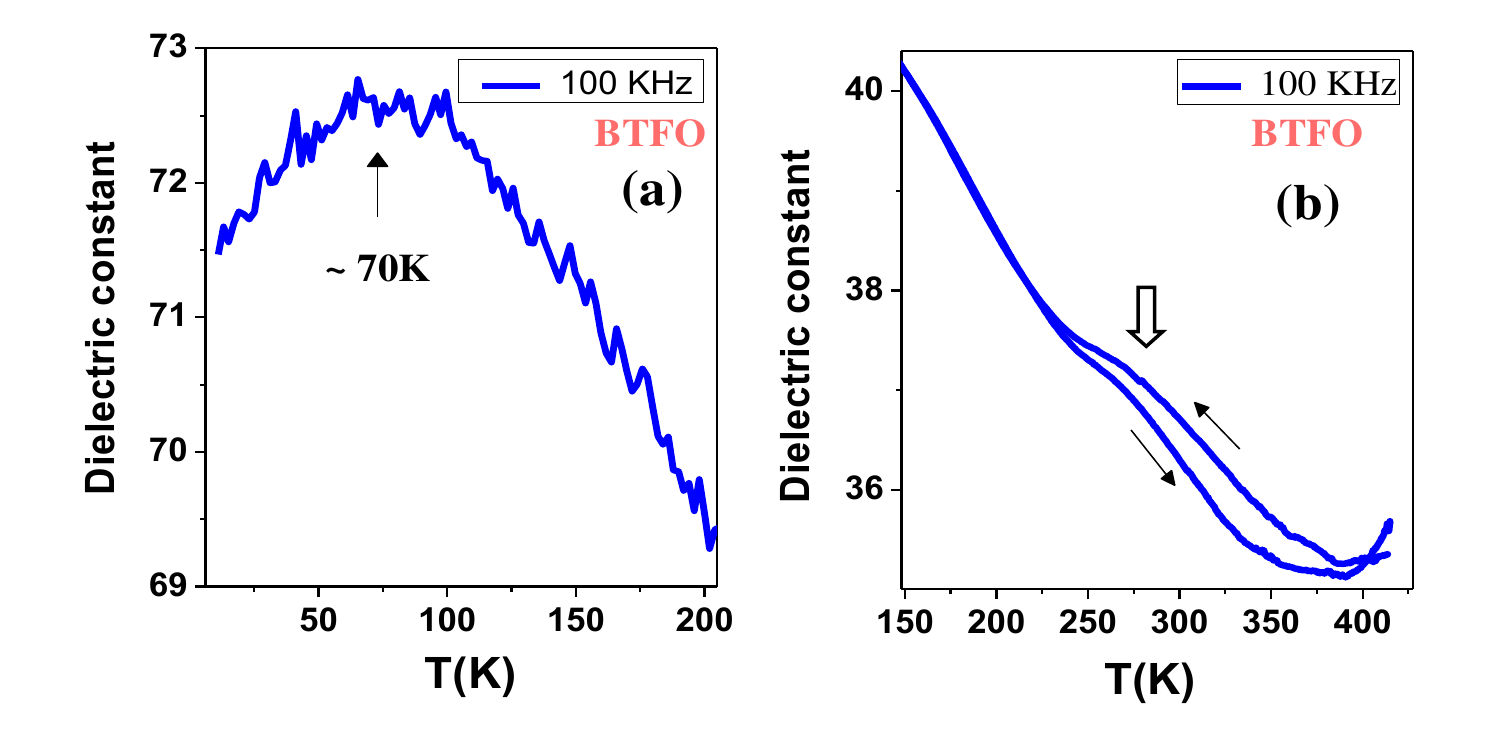}} 
	\caption{(a) and (b) show dielectric anomalies in BTFO around the similar temperature regions as those observed in heat capacity measurements (see the inset to Fig.\ref{Magnetic}(b)), which are indicative to structural transitions.}\label{BTFO-DC}
	
\end{figure*}

	\subsection{Variation of dielectric constant with temperature of pure BaTi$_{0.9}$Fe$_{0.1}$O$_3$  (see Fig.\ref{BTFO-DC})}


\begin{figure*} 
	
	\centering \scalebox{1.2}{\includegraphics{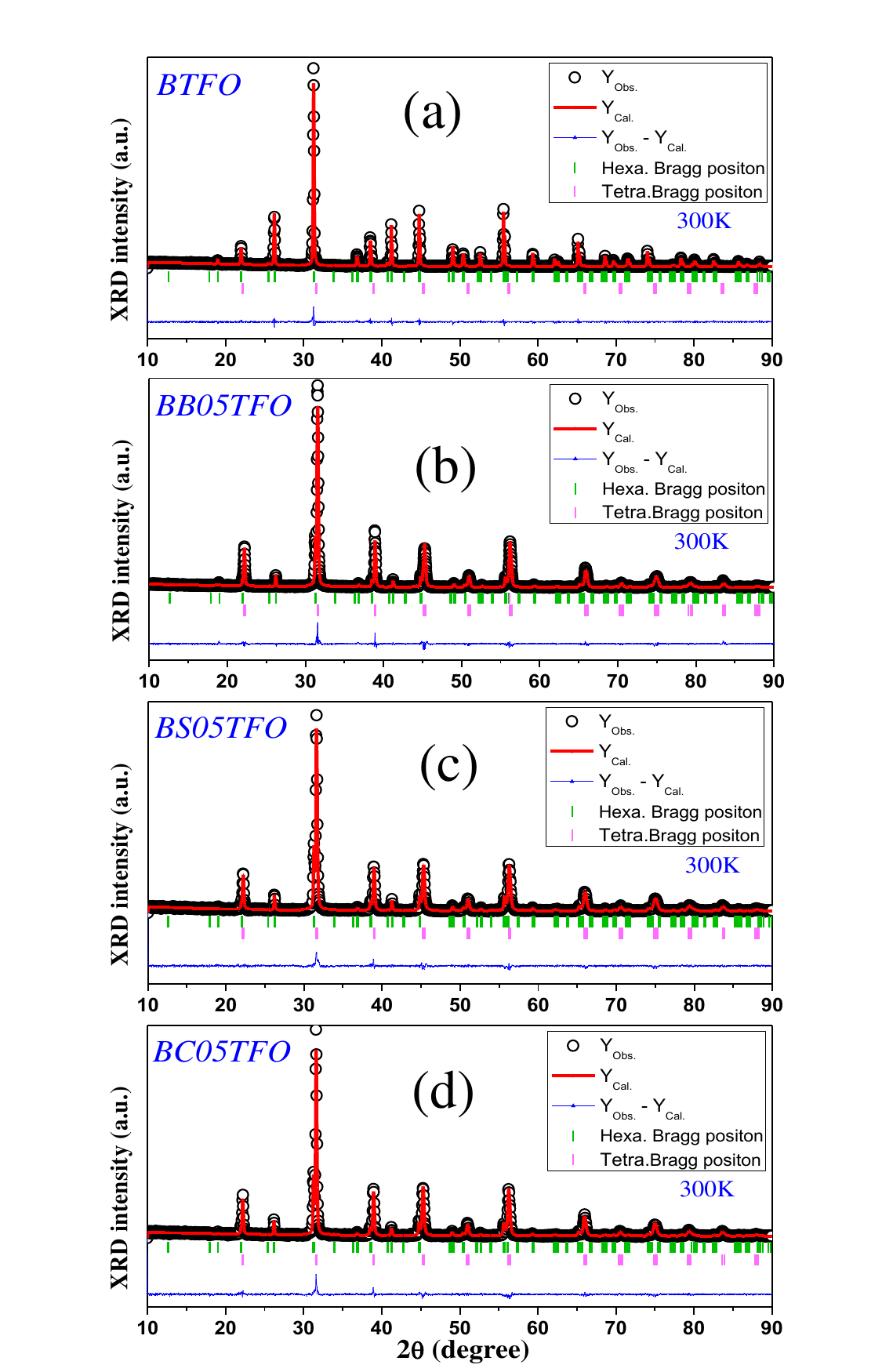}} 
	\caption{Rietveld refinement of room-temperature XRD spectra of (a) BTFO (BaTi$_{0.9}$Fe$_{0.1}$O$_3$), (b) BB05TFO  (Ba$_{0.95}$Bi$_{0.05}$Ti$_{0.9}$Fe$_{0.1}$O$_3$),  (c) BS05TFO (Ba$_{0.95}$Sr$_{0.05}$Ti$_{0.9}$Fe$_{0.1}$O$_3$) and (d) BC05TFO (Ba$_{0.95}$Ca$_{0.05}$Ti$_{0.9}$Fe$_{0.1}$O$_3$) incorporating both hexagonal ({\it{P6$_3$/mmc }}) and tetragonal ({\it{P4mm}}) BaTiO$_3$ phases. }\label{AllRefinement}
	
\end{figure*}

\subsection{Full Rietveld refinement of room temperature XRD spectra  (see Fig.\ref{AllRefinement})}


\begin{figure*} 
	
	\centering \scalebox{0.45}{\includegraphics{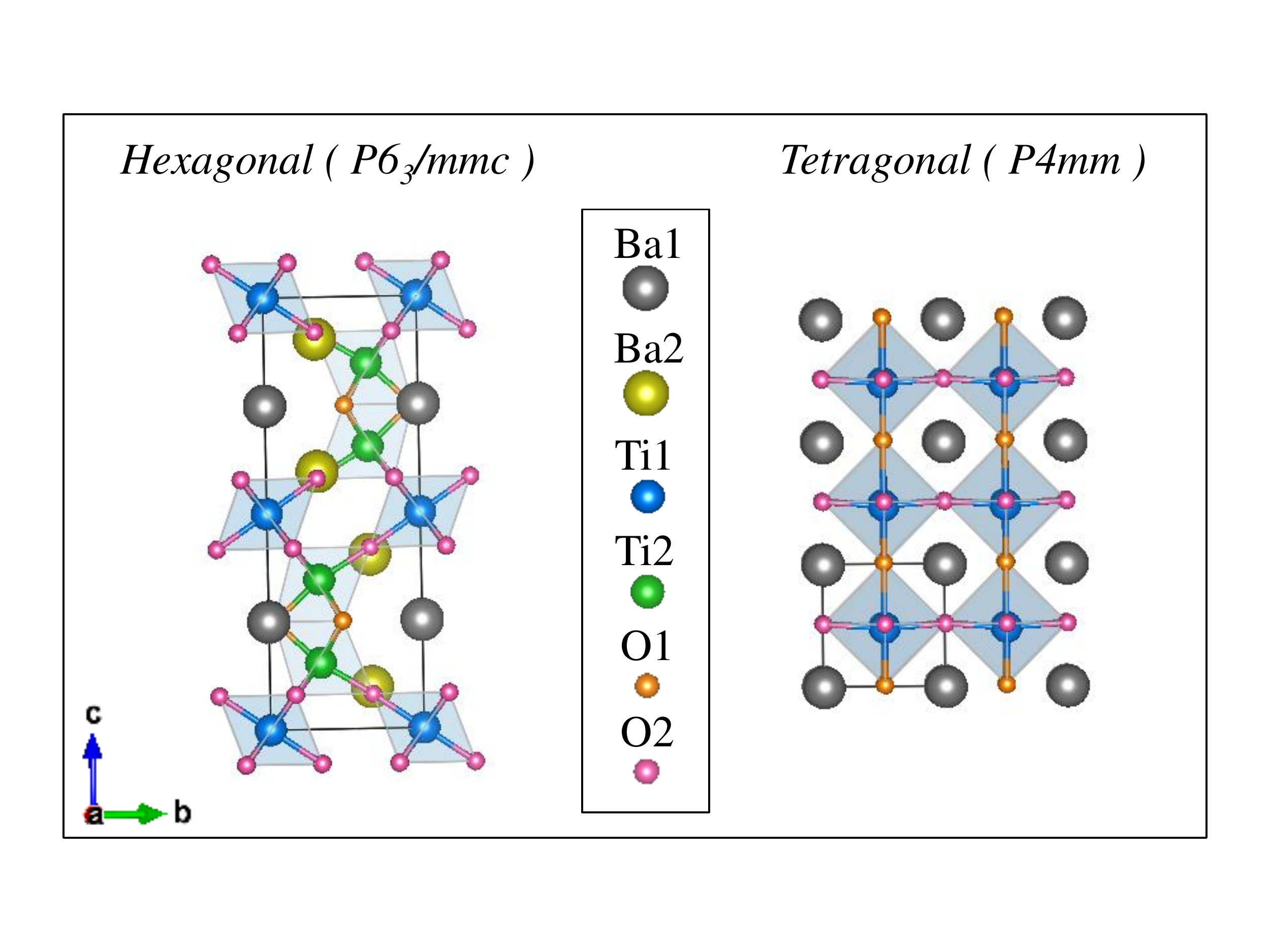}} 
	\caption{Implemented room-temperature crystal structures of hexagonal BaTiO$_3$ and tetragonal BaTiO$_3$ for Rietveld refinement. }\label{Structure}
	
\end{figure*}

\subsection{Room-temperature crystal structures of the hexagonal and tetragonal BTO phases as employed for refinement (see Fig.\ref{Structure})}


\begin{figure*} 
	
	\centering \scalebox{0.6}{\includegraphics{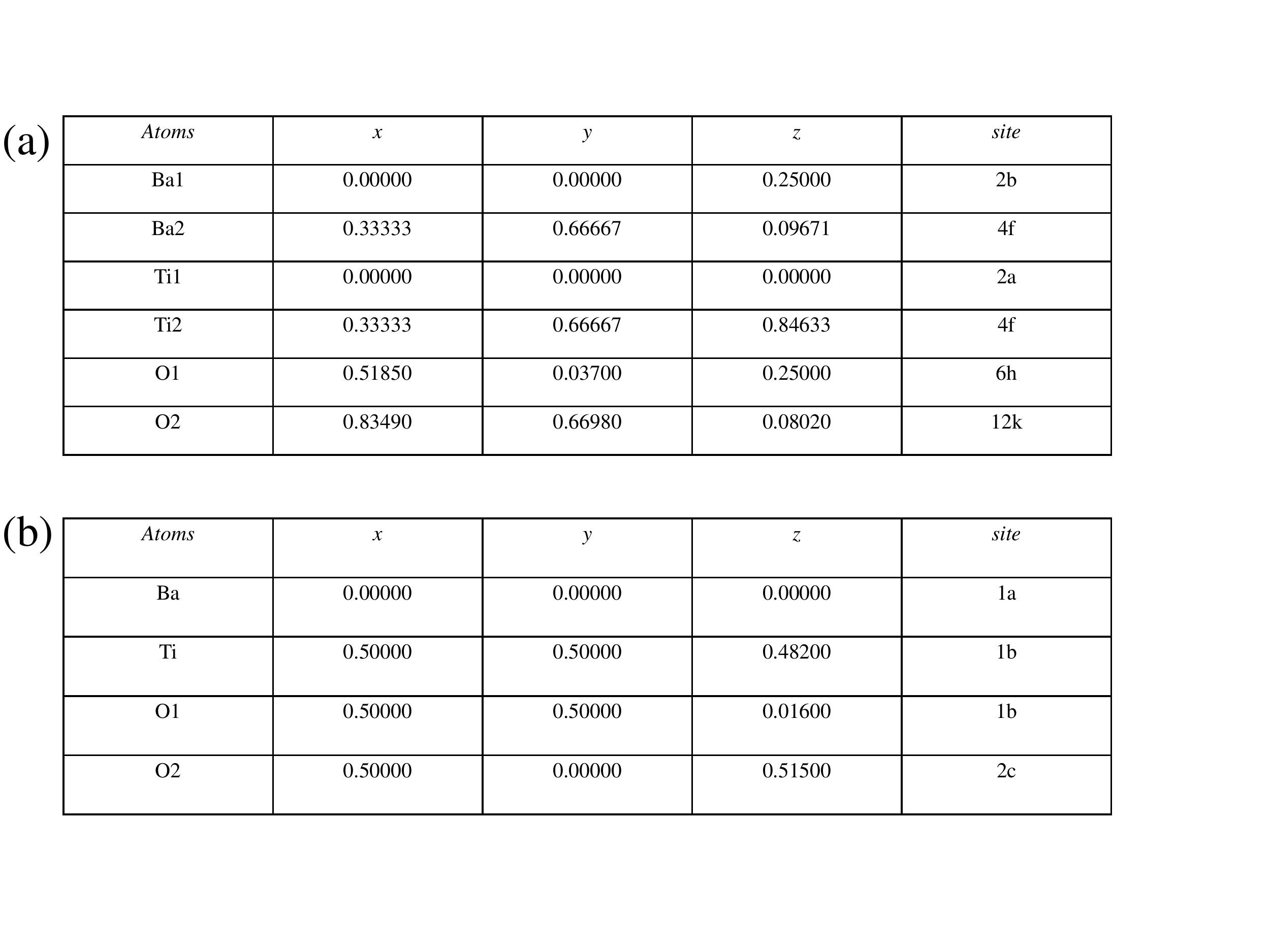}} 
	\caption{(a) and (b) display informations of atomic positions of the adopted crystal structures of hexagonal ({\it{P6$_3$/mmc }}) and tetragonal ({\it{P4mm}}) BaTiO$_3$ phases.}\label{StructuralInformation}
	
\end{figure*}

\subsection{Details of atomic positions of the adopted hexagonal and tetragonal BTO phases  (see Fig.\ref{StructuralInformation})}


\begin{figure*} 
	
	\centering \scalebox{0.75}{\includegraphics{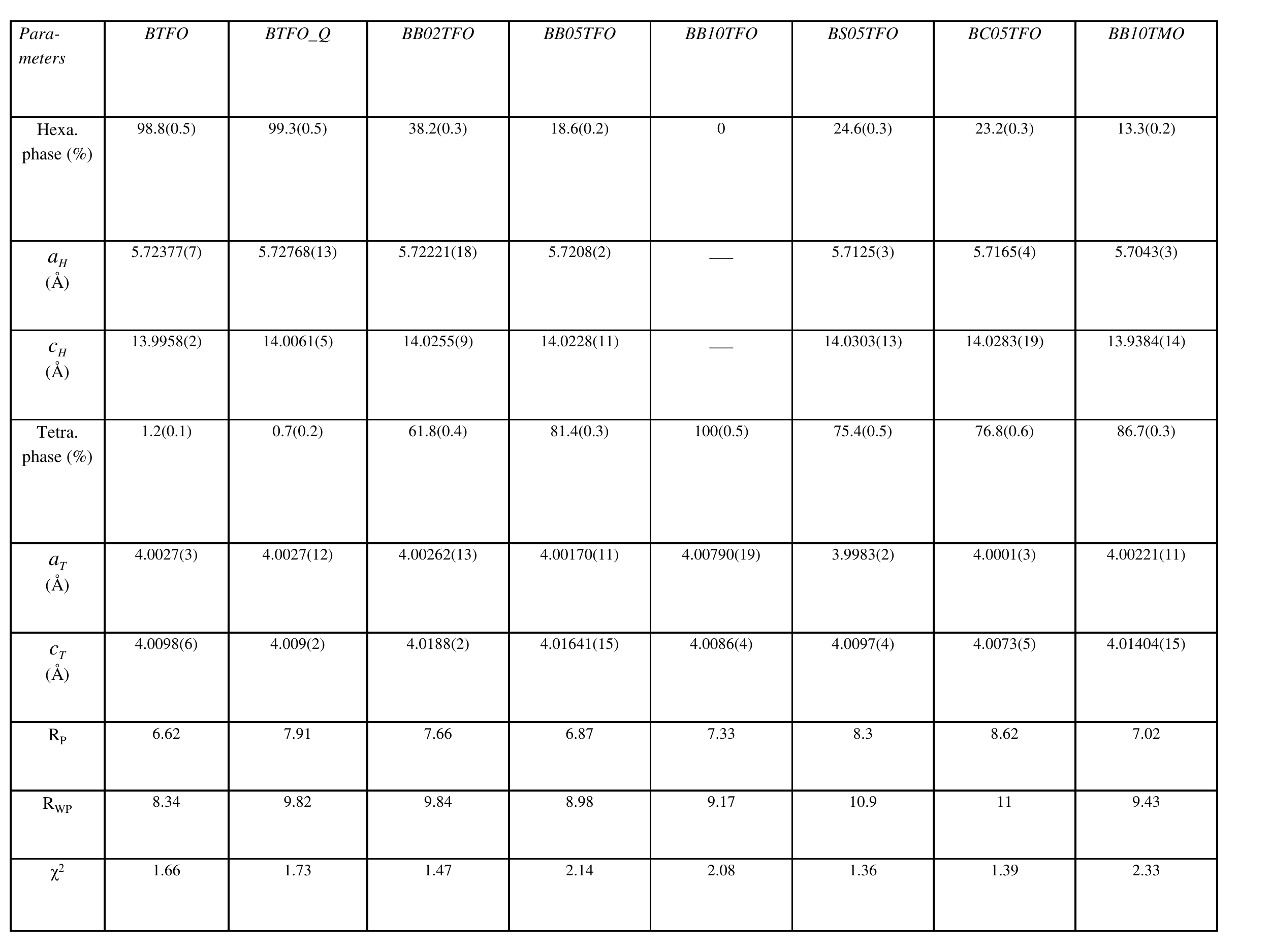}} 
	\caption{Structural parameters obtained from Rietveld refinement.}\label{Refinement details}
	
\end{figure*}

\subsection{Details of structural parameters as obtained from Rietveld refinement  (see Fig.\ref{Refinement details})}


\begin{figure*} 
	
	\centering \scalebox{1}{\includegraphics{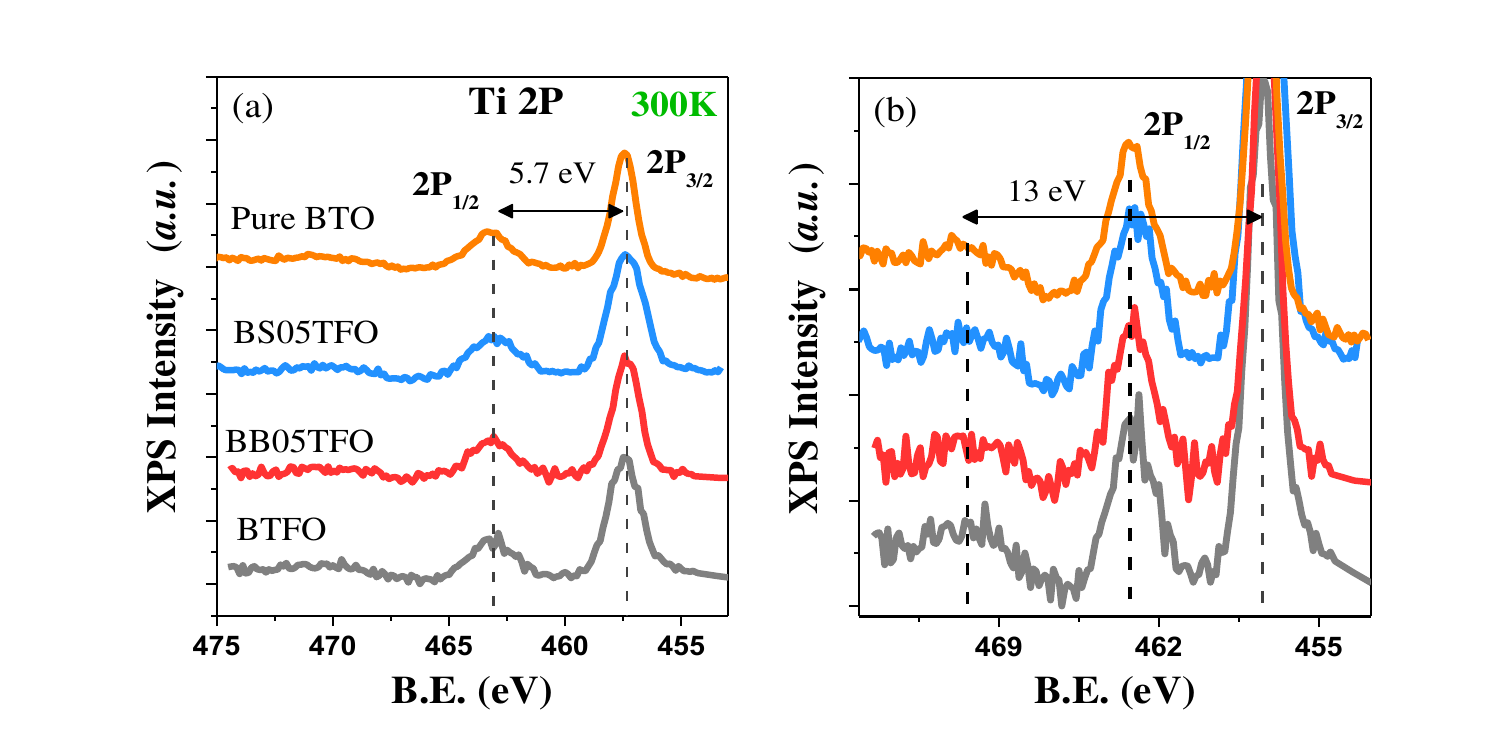}} 
	\caption{(a) Room temperature Ti {\it{2p}} core level XPS spectra of BTFO (BaTi$_{0.9}$Fe$_{0.1}$O$_3$), BB05TFO (Ba$_{0.95}$Bi$_{0.05}$Ti$_{0.9}$Fe$_{0.1}$O$_3$),  BSTFO (Ba$_{0.95}$Sr$_{0.05}$Ti$_{0.9}$Fe$_{0.1}$O$_3$) and BTO (BaTiO$_3$). Here, Ti {\it{2p}} XPS core level spectra show that all these compounds have same spin-orbit splitting of 5.7 eV between Ti 2p$_{3/2}$ and 2p$_{1/2}$ peaks, which is expected for a compound with Ti$^{4+}$  species. Also, all these compounds show weak  satellite features, 13 eV away from the Ti 2p$_{3/2}$ peak, which is also a characteristic of the Ti$^{4+}$ state. }\label{Ti2PXPS}
	
\end{figure*}

\subsection{Room temperature XPS spectra of Ti {\it{2p}} core level  (see Fig.\ref{Ti2PXPS})}


\begin{figure*} 
	
	\centering \scalebox{1}{\includegraphics{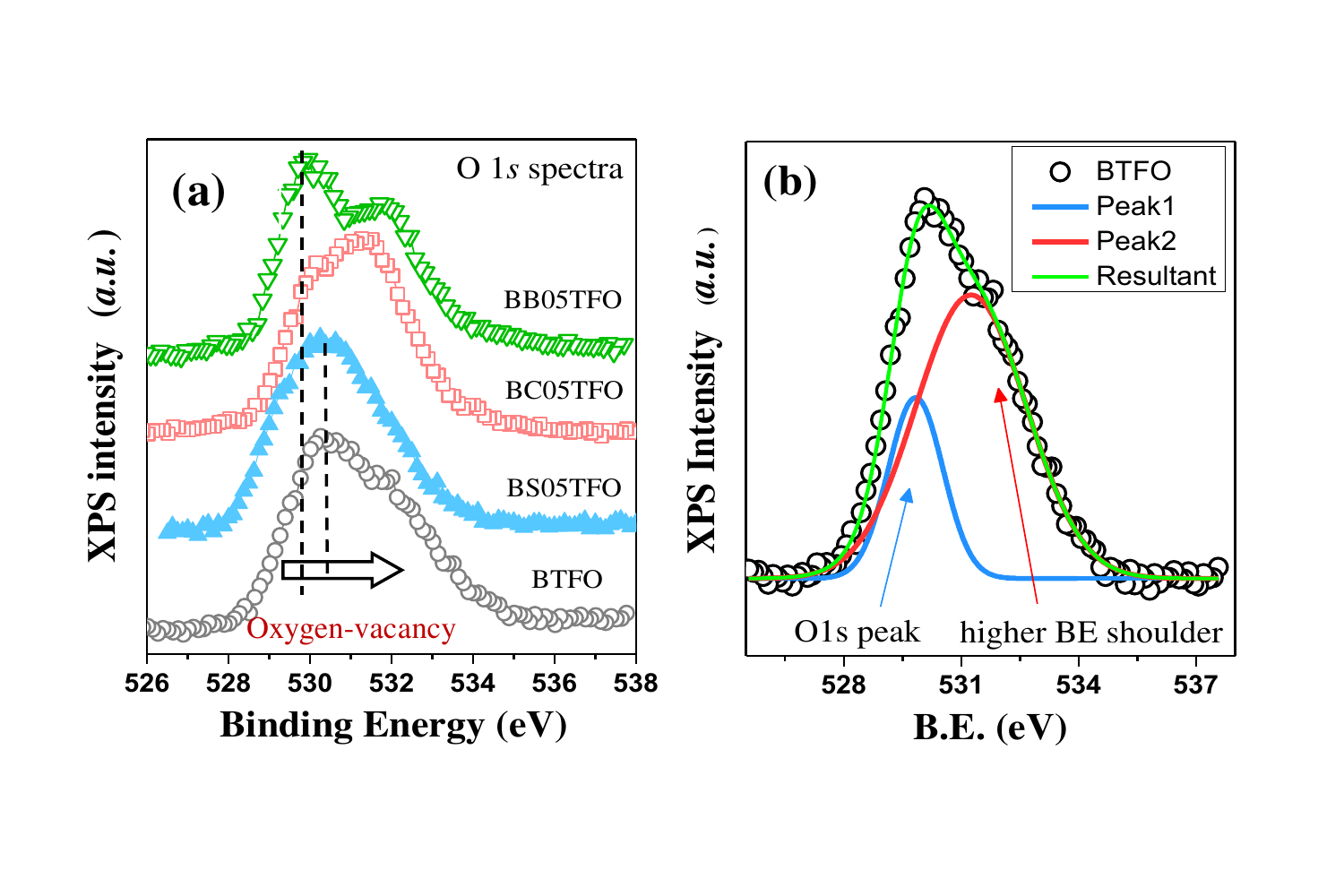}} 
	\caption{(a) displays room temperature O{\it{1s}} XPS spectra of BTFO (BaTi$_{0.9}$Fe$_{0.1}$O$_3$), BB05TFO  (Ba$_{0.95}$Bi$_{0.05}$Ti$_{0.9}$Fe$_{0.1}$O$_3$),  BC05TFO (Ba$_{0.95}$Ca$_{0.05}$Ti$_{0.9}$Fe$_{0.1}$O$_3$)  and  BS05TFO (Ba$_{0.95}$Sr$_{0.05}$Ti$_{0.9}$Fe$_{0.1}$O$_3$), and  (b) shows two peak fitting of BTFO O{\it{1s}} spectrum, where the higher binding-energy shoulder arises from defect sites like oxygen-vacancies and adsorbed hydrocarbons. Here, O 1s XPS  spectra of  BTFO, BS05TFO, BC05TFO and BB05TFO show that all these compounds are associated with some amount of oxygen vacancies which can be correlated with the appearance of higher binding-energy shoulder in O-1s spectra. }\label{Oxygenvacancy}
	
\end{figure*}

\subsection{Room temperature O{\it{1s}} XPS spectra  (see Fig.\ref{Oxygenvacancy})}


\end{document}